\documentclass[useAMS,usenatbib]{mnras}

\usepackage{mathptmx}
\usepackage{ amssymb }
\usepackage{multirow}
\usepackage{psfrag}
\usepackage{siunitx}
\usepackage{pspicture}
\usepackage{graphicx}
\usepackage{amsmath}
\usepackage{color}
\usepackage{url}
\usepackage{threeparttable}
\usepackage{caption}
\usepackage{lscape}
\usepackage{booktabs}
\usepackage{subcaption}
\usepackage{grffile}
\usepackage{multirow}
\usepackage{multicol}
\usepackage{dcolumn}
\usepackage{hyperref}
 
\usepackage{setspace}
\usepackage{xspace}
\newcolumntype{P}[1]{>{\centering\arraybackslash}p{#1}}

\newcommand{\C}{C{\sc iii}]\xspace} 
\newcommand{\CC}{C{\sc ii}]\xspace}
\newcommand{\CIV}{C{\sc iv}\xspace}

\pubyear{2017}

\title[{C{\sc ii}], C{\sc iii}] and C{\sc iv} emitters at $z\sim0.7-1.5$}]{A ${\bf 1.4}$ deg${\bf ^2}$ blind survey for C{\sc ii}], C{\sc iii}] and C{\sc iv} at ${\bf z\sim0.7-1.5}$. I: \\
nature, morphologies and equivalent widths}
\author[A. Stroe et al.]{Andra Stroe$^{1}$\thanks{E-mail: astroe@eso.org}\thanks{ESO Fellow}, David Sobral$^{2,3}$, Jorryt Matthee$^{2}$, Jo\~ao Calhau$^{3}$, Ivan Oteo$^{4,1}$\\
$^{1}$European Southern Observatory, Karl-Schwarzschild-Str. 2, 85748, Garching, Germany\\
$^{2}$Leiden Observatory, Leiden University, P.O.\ Box 9513, NL-2300 RA Leiden, The Netherlands\\
$^{3}$Department of Physics, Lancaster University, Lancaster, LA1 4YB, UK\\
$^{4}$Institute for Astronomy, University of Edinburgh, Royal Observatory, Blackford Hill, Edinburgh EH9 3HJ UK}
\begin{document}

\maketitle
\begin{abstract}
{While traditionally associated with active galactic nuclei (AGN), the properties of the \CC ($\lambda=2326$\,{\AA}), \C ($\lambda,\lambda=1907,1909$\,{\AA}) and \CIV ($\lambda,\lambda=1549, 1551$\,{\AA}) emission lines are still uncertain as large, unbiased samples of sources are scarce. We present the first blind, statistical study of \CC, \C and \CIV emitters at $z\sim0.68,1.05,1.53$, respectively, uniformly selected down to a flux limit of \mbox{$\sim4\times10^{-17}$~erg\,s$^{-1}$\,cm$^{-1}$} through a narrow band survey covering an area of $\sim1.4$ deg$^2$ over COSMOS and UDS. We detect 16 \CC, 35 \C and 17 \CIV emitters, whose nature we investigate using optical colours as well as {\it HST}, X-ray, radio and far infra-red data. We find that $z\sim0.7$ \CC emitters are consistent with a mixture of blue (UV slope $\beta=-2.0\pm0.4$) star forming galaxies with disky {\it HST} structure and AGN with Seyfert-like morphologies. Bright \CC emitters have individual X-ray detections as well as high average black hole accretion rates (BHAR) of $\sim0.1$ $M_{\odot}$\,yr$^{-1}$. \C emitters at $z\sim1.05$ trace a general population of SF galaxies, with $\beta=-0.8\pm1.1$, a variety of optical morphologies, including isolated and interacting galaxies and low BHAR ($<0.02$ $M_{\odot}$\,yr$^{-1}$). Our \CIV emitters at $z\sim1.5$ are consistent with young, blue quasars ($\beta\sim-1.9$) with point-like optical morphologies, bright X-ray counterparts and large BHAR ($0.8$ $M_{\odot}$\,yr$^{-1}$). We also find some surprising \CC, \C and \CIV emitters with rest-frame equivalent widths which could be as large as $50-100$\,{\AA}. AGN or spatial offsets between the UV continuum stellar disk and the line emitting regions may explain the large EW. These bright \CC, \C and \CIV emitters are ideal candidates for spectroscopic follow up to fully unveil their nature.}
\end{abstract}

\section{Introduction}\label{sec:intro}

Rest-frame ultra-violet (UV) emission lines are of great importance for extragalactic astrophysics as they can be used to infer gas metallicities, temperatures and the strength of the ionising field \citep[e.g.][]{2003ApJ...588...65S, 2006agna.book.....O}. Observing these lines in the local Universe is challenging. However, for galaxies at large cosmic distances the rest-frame UV emission lines are redshifted into the easily observable optical range. Historically rest-frame UV spectra were mainly used to spectroscopically confirm UV-bright galaxies at $z\gtrsim3$ selected through the Lyman Break technique \citep[Lyman Break galaxies, LBG,][]{1996ApJ...462L..17S, 1997ApJ...481..673L}. In the last 15 years, strong rest-frame UV lines have been discovered in a variety of galaxies at $z>1$, which could be used to constrain the physics in high-redshift counterparts of local galaxies \citep{2000MNRAS.311...23B, 2003ApJ...588...65S,2010ApJ...719.1168E, 2015ApJ...814L...6R}. 

\renewcommand{\arraystretch}{1.2}
\begin{table*}
\begin{center}
\caption{Line emitters studied in this work, with rest-frame wavelength, ionisation energy $\chi$ \citep{2002ASPC..284..111V}, luminosity distance $D_{\rm L}$ and redshift range traced by the NB392 filter, with central wavelength $3919$\,{\AA} and FWHM of $52$\,{\AA}. The sample is drawn from the emission line catalogue presented in \citet{Sobral2017}, which focuses on selection and properties of Ly$\alpha$ emitters at $z\sim2.2$.}
\begin{tabular}{l c c c c c}
\hline\hline
Line & $\lambda_\mathrm{line}$ & $\chi$ & $z_\mathrm{line}$ & $D_{\rm L}$& Comments\\ 
     & (\AA)                   & (eV)   & at FWHM           & ($10^3$ Mpc) & \\ \hline
\CC & 2326 & 11.3 & $0.673-0.696$ & 4.14 & likely traces shocks around AGN \\
\C & 1907, 1909 & 24.4 & $1.039-1.066$ & 7.04 & produced in SF and BLR of AGN \\ 
\CIV & 1549, 1551 & 47.9 &   $1.513-1.546$ & 11.47 & likely produced in BLR of AGN or in gas around very massive stars \\
\hline
\end{tabular}
\label{tab:lines}
\end{center}
\end{table*}
\renewcommand{\arraystretch}{1.1}

Intrinsically the brightest rest-frame UV line is Ly$\alpha$, which is produced in H{\sc II} regions as well as in galaxies with an active galactic nucleus (AGN) \citep[e.g.][]{2008ApJS..176..301O, 2009A&A...498...13N, 2010ApJ...711..928C, 2015PASA...32...27H,2016MNRAS.458..449M, Matthee2017, Sobral2017}. However, it is difficult to interpret observations of Ly$\alpha$, as the line is scattered by neutral hydrogen and easily absorbed by dust \citep[e.g.][]{2014PASA...31...40D, 2015PASA...32...27H}. The amount of Ly$\alpha$ that escapes a galaxy therefore depends on the properties of the interstellar medium. 

Other UV lines, such as \CC ($\lambda=2326$\,{\AA}), \C ($\lambda,\lambda=1907,1909$\,{\AA}) and \CIV ($\lambda,\lambda=1549, 1551$\,{\AA})  (Table~\ref{tab:lines}) have recently been explored as they can be relatively bright compared to Ly$\alpha$ and can be used either individually or in combination to constrain the physics of the host galaxy. \C \citep[ionisation potential of 24.4 eV, ][]{2002ASPC..284..111V}, for example, was found to be the strongest UV line after Ly$\alpha$ in stacks of LBGs at $z\sim3$ \citep{2003ApJ...588...65S} and about 10 per cent of the observed strength of Ly$\alpha$ in faint, lensed galaxies at $1.5<z<3$ \citep{2014MNRAS.445.3200S}. \CIV (47.9 eV) is also a bright UV line. From photoionisation modelling, the \CIV ratio to \C ranges from 0.1 for low metallicity star-forming ($\lesssim0.3 Z_\odot$), up to 100 for high, supersolar metallicity ($\sim 2-2.5 Z_\odot$). The typical \CIV to \C ratio is $\sim1$ for star forming galaxies of solar-type metallicity \citep[e.g.][]{2016MNRAS.462.1757G,2016MNRAS.456.3354F}.

High ionisation rest-frame UV emission lines of Carbon were originally thought to originate from close to the AGN engine as they require a strong radiation field and high temperatures. \C is a high-ionisation, intercombination doublet ($1907, 1909$\,{\AA}) expected to be mostly produced in the outer parts of the broad line region (BLR) of the AGN \citep{2006agna.book.....O}. However, at $z\sim2-3$, \C emitters are also found in star forming (SF) galaxies and trace a slightly sub-solar metallicity, a high ionization parameter and a hard radiation field \citep{2015ApJ...814L...6R, 2014ApJ...790..144B}. Photoionisation models presented in \citet{2016ApJ...833..136J} indicate that \C can be produced in starburst galaxies and is the strongest line (with $\lambda<2700$\,{\AA}) after Ly$\alpha$. 

UV collisionally excited lines such as \CC and \CIV are stronger in very high temperature regions ($2\times10^4-10^5$\,K) in the cooling region behind shocks around the AGN than in areas with lower temperature such as those that can be reached with photoionisation \citep[$10^4$\,K,][]{2006agna.book.....O, 1998ApJ...493..571A}. \CC and \CIV are therefore expected to be more strongly produced in AGN hosting galaxies. Indeed, \citet{1977MNRAS.178P..67B} found a correlation between the strength of the resonant \CIV ($1548, 1551$\,{\AA}) line and the continuum luminosity, indicating that the line emitting gas is located very close to the ionisation source. While traditionally associated with BLR emission, \CIV was found to be correlated with gas temperature and an intense radiation field \citep{2006agna.book.....O}, such as the one caused by AGN or by massive stars after a recent SF episode \citep{2014MNRAS.445.3200S, 2017ApJ...836L..14M, 2017ApJ...839...17S}. In a study of radio galaxies, \citet{2000A&A...362..519D} noted that \CC is five times more sensitive to shock ionisation than high ionisation UV lines, such as \CIV. Therefore, there is compelling evidence that the semi-forbidden \CC \citep[11.3 eV, ][]{2002ASPC..284..111V} line traces shocks and in combination with other lines is effective in determining the power source of ionisation \citep{2000MNRAS.311...23B, 2002RMxAC..13..155B}. 

While independently \CC, \C and \CIV trace gas metallicity and electron density, in combination they can be used as estimators which are, to first order, independent of abundance, metallicity and dust-extinction \citep{2002RMxAC..13..155B}. Therefore, \CC, \C and \CIV line ratios are some of the best diagnostics to separate excitation by fast shocks and photoionisation in a hard photon spectrum  \citep{1998ApJ...493..571A}.

Given their relative strength to Ly$\alpha$, \C and \CIV have been proposed as a good avenue for spectroscopically confirming high redshift galaxies \citep[][]{2014MNRAS.445.3200S, 2015MNRAS.450.1846S, 2015MNRAS.454.1393S}, particularly within the epoch of reionisation, when Ly$\alpha$ scattering is expected to increase leading to a significant decrease of surface brightness. As a result, in recent years, targeted searches for \C and \CIV emitters at high redshift have emerged. For example, \citet{2015ApJ...814L...6R} detect \C in 11 $z\sim1.6-3$ lensed galaxies, and \citet{2014ApJ...790..144B} find strong \C in a $z\sim3.63$ lensed starburst. \citet{2015MNRAS.450.1846S} present tentative detections of \C in two galaxies at $z>6$, while \citet{2016arXiv161200902D} detect \C in one galaxy at $z\sim5.7$. By contrast, \citet{2015ApJ...805L...7Z} do not obtain a detection of \C in a sample of 7 $z\sim7-8$ photometric candidates. Very recently, \citet{2017ApJ...838...63D} presented a spectroscopic study of a sample of continuum selected \C emitters at $z\sim1$. With a detection rate of $\sim20$ per cent, their \C emitters have much lower EW ($1.3$\,{\AA}) than higher redshift examples. \citet{2017ApJ...838...63D} also found that the stronger EW sources appear in fainter, bluer and lower-mass galaxies. \citet{2017ApJ...839...17S} and \citet{2017ApJ...836L..14M} obtain a detection of \CIV in a multiply-lensed $z=6.1$, Ly$\alpha$-emitting SF galaxy, but without a \C detection. Despite the growing number of detections, the samples of \C and \CIV emitters suffer from selection biases (e.g. spectroscopically selected, lensed sources, redshift known from Ly$\alpha$). 

Despite the potential importance of \CC, \C and \CIV for understanding AGN physics and the nature of stellar populations at high redshift, not much is known about these emitters in a statistical sense, as no blind studies have been performed. As such, the nature, number densities and EW distributions are largely unknown. We seek to improve our understanding of \CC, \C and \CIV emitters by performing the first blind survey of these lines, without any preselection in terms of Ly$\alpha$ or UV properties. Our sample is uniformly selected down to a flux of $\sim4\times10^{-17}$ erg\,s$^{-1}$\,cm$^{-2}$, in three redshift slices around $z\sim0.7$, $\sim1.0$, $\sim1.5$ for \CC, \C and \CIV, respectively. The limiting observed EW is 16\,{\AA} and the limiting $U$ magnitude is $\sim26.5$. The sources were discovered by exploring the $\sim1.4$ deg$^2$ CAlibrating LYMan-$\alpha$ with H$\alpha$ NB survey \citep[CALYMHA,][]{2016MNRAS.458..449M, Sobral2017} over the COSMOS and UDS fields. 

Our results are presented in two parts. In this paper (Paper I), we use the emission line data in combination with multiwavelength observations in the optical, radio, X-ray and far infra red (FIR) to unveil the characteristics of individual \CC, \C and \CIV emitters selected with the CALYMHA survey. We study the likely physical origin of the emission lines and how their properties compare with AGN and SF galaxies at similar redshifts. In the companion paper \citep[Paper II, ][]{PaperII} we investigate the statistical properties of the \CC, \C and \CIV emitters through luminosity functions (LF) and obtain the volume-average line ratios relative to e.g. Ly$\alpha$ and H$\alpha$.

We organise the paper as follows: in Section \ref{sec:data} we present the CALYMHA parent sample, while in Section~\ref{sec:select} we select the \CC, \C and \CIV emitters. We discuss the colour and EW properties of the emitters as well as their Hubble Space Telescope ({\it HST}), radio, far infra-red (FIR) and X-ray properties in Section~\ref{sec:properties}. The interpretation of our results and the implication for the physics \CC, \C and \CIV production can be found in Section~\ref{sec:discussion}, with conclusions and outlook in Section \ref{sec:conclusion}.

Throughout the paper, we use a flat $\Lambda$CDM cosmology ($H_{0}=70$\,km\,s$^{-1}$\,Mpc$^{-1}$, $\Omega_M=0.3$, $\Omega_{\Lambda}=0.7$), and perform calculations with the aid of the \citet{2006PASP..118.1711W} cosmology calculator. All magnitudes are in the AB system and we use a \citet{2003PASP..115..763C} initial mass function (IMF).

\begin{figure}
\centering
\includegraphics[trim=0cm 0cm 0cm 0cm, width=0.489\textwidth]{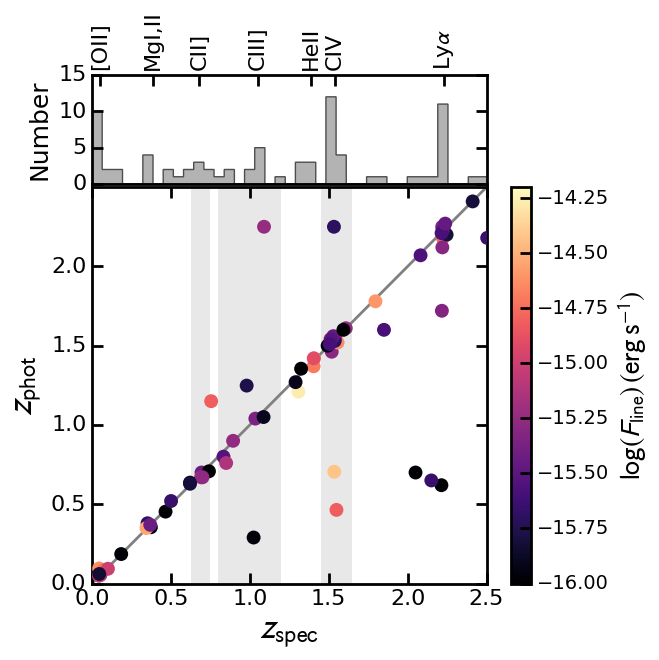}
\caption{Photometric versus spectroscopic redshift for our NB392 line emitters, using data from e.g. \citet{2009ApJ...690.1236I}, \citet{2010MNRAS.401.1166C} and \citet{Lilly09}, as well as our own X-SHOOTER Very Large Telescope follow-up. The grey shaded areas indicate the redshift ranges where the NB filter is sensitive to \CC, \C and \CIV. The top panel shows the distribution of spectroscopic redshifts, where we also mark the main emission lines picked up by the NB filter.}
\label{fig:z}
\end{figure}

\section{Survey description}\label{sec:data}

We use the CALYMHA sample of emission line galaxies to select \CC ($z\sim0.63$), \C ($z\sim1.05$) and \CIV ($z\sim1.53$) line emitters in the COSMOS and UDS fields. The CALYMHA survey design, observations and data reduction is presented in full in \citet{Sobral2017} and here we give a brief summary of the survey strategy and goals. The programme surveyed a combined area of $\sim1.4$ deg$^2$ in the COSMOS and UDS fields using a narrow band (NB) filter (NB392, central wavelength $\lambda_C = 3918$\,{\AA} and width $\Delta \lambda = 52$\,{\AA}) mounted on the Isaac Newton Telescope\footnote{\url{http://www.ing.iac.es/Astronomy/telescopes/int/}}. In combination with ancillary broad band (BB) $U$ data ($\lambda_C = 3750$\,{\AA}, $\Delta \lambda = 720$\,{\AA}), the NB filter was designed to select line emitters with a particular focus on Ly$\alpha$ emitters at $z\sim2.23$, and cross-match them with H$\alpha$ galaxies at the same redshift \citep{2013MNRAS.428.1128S}. The main goal of the survey is to unveil the nature of Ly$\alpha$ emitter by studying the luminosity function (LFs) and determining Ly$\alpha$ escape fractions as function of galaxy properties both for H$\alpha$ and Ly$\alpha$ selected samples at $z\sim2.2$ \citep{2016MNRAS.458..449M, Sobral2017}. 
 
The CALYMHA COSMOS+UDS survey selected a total of 440 line emitters down to a $3\sigma$ line flux limit of $\sim4\times10^{-17}$ erg\,s$^{-1}$\,cm$^{-2}$, down to an observed EW limit of 16\,{\AA}. Based on spectroscopic and photometric redshifts, the emitter population contains a significant fraction of \CC, \C and \CIV emitters \citep{Sobral2017}, thus rendering CALYMHA an ideal sample to study these emitters in a statistical, unbiased way with a clear selection function. Given the width of the NB filter and their rest-frame wavelength, the line emitters are traced over a narrow redshift range (see Table~\ref{tab:lines}).
\begin{figure}
\centering
\includegraphics[trim=0cm 0cm 0cm 0cm, width=0.489\textwidth]{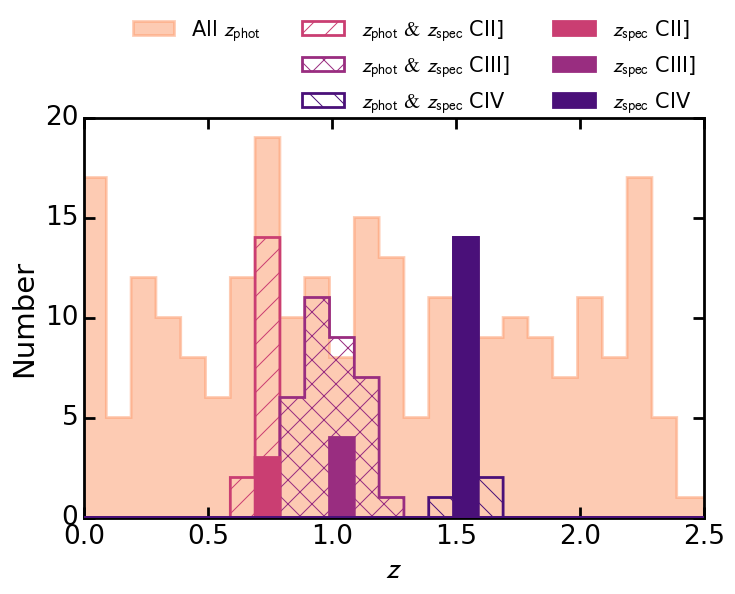}
\caption{Histogram of all photometric redshifts, focusing on \CC, \C and \CIV. Note the narrow ranges chosen for selection of sources based on photometric redshifts, ensuring that low redshift ($z < 0.4$) emitters such as [O{\sc ii}], [Ne{\sc v}], Mg{\sc i} and Mg{\sc ii} are rejected.}
\label{fig:zhist}
\end{figure}

\subsection{Ancillary data}
In addition to the CALYMHA NB and {\it U} band data, we use ancillary spectroscopic and photometric redshifts and photometry from the COSMOS and UDS surveys \citep{2007ApJS..172...99C, 2007MNRAS.379.1599L, 2009ApJ...690.1236I,2016ApJS..224...24L}. About $40$ per cent of our sources are faint in the $i$ and $K$ bands or are located in masked regions and are thus not included in the publicly available COSMOS and UDS catalogues. There are 80 emitters with spectroscopic redshifts, most of which also have a photometric redshift \citep[Fig.~\ref{fig:z}, data from][]{Yamada05, Simpson06, van_breu07, Geach008, 2008ApJS..176..301O,Smail08,Lilly09, Ono09}. However, in 10 cases, only a spectroscopic redshift is available. We also include the redshifts derived by \citet{Sobral2017} from dual, triple and quadruple detection of emission lines in NB filters. These very precise photometric redshifts have accuracies close to a spectroscopic measurement. The total tally for sources with redshifts (spectroscopic or photometric) is 269, or 61 per cent of the total number of emitters. 

\begin{table*}
\begin{center}
\caption{Criteria for selecting a source as a \CC, \C or \CIV emitter. The $z_{\rm spec}$ ranges used correspond to the full transmission range covered by the NB filter. Note that we are using conservative $z_{\rm phot}$ cuts to minimise contamination. Sources selected as Ly$\alpha$ by \citet{Sobral2017} using colour-colour selections were removed from the sample. The number of sources of each type, selected based on $z_{\rm spec}$ and additional $z_{\rm phot}$ are also listed.}
\begin{tabular}{l c c c c c c}
\hline\hline
Line & $z_{\rm spec}$ selection range & $z_{\rm spec}$ sources   & $z_{\rm phot}$ selection range & $z_{\rm phot}$ sources & All \\ 
     &           &                & (sources without $z_{\rm spec}$)                      &  &  \\ \hline
\CC &  $0.661-0.707$ & \phantom{0}3 & $0.63-0.75$ & 13 & 16  \\
\C & $1.025-1.080$ &  \phantom{0}4 & $0.8-1.2$ & 30 & 34  \\ 
\CIV &  $1.486-1.563$ & 14 & $1.4-1.7$ & \phantom{0}3  & 17 \\
\hline
\end{tabular}
\label{tab:numbers}
\end{center}
\end{table*}

We also explore the deep, publicly available \textit{HST} data in the F814W filter \citep{2007ApJS..172..196K, 2010MNRAS.401..371M}, \textit{Chandra} space telescope X-ray observations \citep{2009ApJS..184..158E}, FIR \textit{Herschel} data \citep{2012MNRAS.424.1614O} and radio Very Large Array images at $1.4$\,GHz \citep{2004AJ....128.1974S, 2010ApJS..188..384S} in the COSMOS field to further investigate the nature of the line emitters. We employ direct detections as well as stacking techniques for this purpose. We note that the Chandra deep data is only available in a sub-area of the COSMOS field, hence only a fraction of the sources will have counterparts and/or coverage. The UDS field is partly covered with {\it HST} data as part of the CANDELS survey \citep{2011ApJS..197...36K}.

\section{Selecting \CC, \C and \CIV emitters}\label{sec:select}

In order to select \CC, \C and \CIV emitters at the redshifts traced by the NB392 filter, we use a combination of spectroscopic and photometric redshifts. 

\subsection{Redshifts}\label{sec:redshifts}

For the COSMOS field, \citet{2009ApJ...690.1236I} derived photometric redshifts using a range of templates, including star, galaxy and quasar templates for an $i$ band selected sample. Blindly using the galaxy templates results in large discrepancies between the chosen photometric redshift and the true redshift, when spectroscopy is available. Keeping in mind that a fraction of \CC, \C and \CIV emitters is possibly tracing AGN activity, we expect in many cases the quasar templates to perform better. \citet{2009ApJ...690.1236I} also provide the $\chi^2$ for the best fit stellar, galaxy and quasar template. We found that simply choosing the template which provided the lowest $\chi^2$ fit worked well: for the sources with both $z_{\rm phot}$ and $z_{\rm spec}$, the two estimates matched (see Fig.~\ref{fig:z}). When choosing the best $z_{\rm phot}$ estimate based on $\chi^2$, $88$ per cent of the photometric redshifts are within 0.1 of the spectroscopic ones. In cases where neither template was a good fit (high $\chi^2>100$), all photometric redshift estimates were catastrophically off. 

We also tested the new COSMOS $z_{\rm phot}$ catalogue presented in \citet{2016ApJS..224...24L} using the same method of selecting the best template (minimising the $\chi^2$), but found that the \citet{2009ApJ...690.1236I} photometric redshifts correlate better with the spectroscopic redshifts in our sample. \citet{2016ApJS..224...24L} is selected in the near infrared and \citet{2009ApJ...690.1236I} in the optical. Since our sources are optically selected (in the very blue), it is maybe unsurprising \citet{2009ApJ...690.1236I} $z_{\rm phot}$ work better, given their weighting towards optical bands. In the case of UDS, a single photometric redshift estimate is available \citep{2010MNRAS.401.1166C}.

Overall, for the entire sample, 84 per cent of photometric redshifts are within 0.1 of the spectroscopic redshift (Fig.~\ref{fig:z}). The sample is however not spectroscopically complete, especially for fainter sources, so the photometric redshift accuracy derived here is not necessarily applicable for all the sources without a spectroscopic redshift (Fig.~\ref{fig:zhist}). 

\subsection{Final selection criteria}
For a source to make the \CC, \C or \CIV emitter selection, we first remove all sources selected as Ly$\alpha$ by \citet{Sobral2017}. It then has to fulfill at least one of the criteria listed below. We summarise the criteria in Table~\ref{tab:numbers} and describe them below: 

\begin{itemize}
\item A spectroscopic redshift within the range probed by the respective filter, within two FWHM. We choose this wider range since the filter transmission drops slowly towards its wavelength edges, effectively being sensitive to emitters at twice the FWHM. This also accounts for broad lines.
\item If spectroscopy is not available, we select a source if is has a photometric redshift within $\sim0.2$ of the redshift range the NB filter is sensitive to. Note our very conservative cuts are chosen to maximise the purity of the sample.
\end{itemize}

The redshift distribution of sources selected as \CC, \C and \CIV is shown Fig.~\ref{fig:zhist}. The narrow photometric redshift ranges chosen for selection ensures that we do not include bright, low redshift emitters such as [O{\sc ii}] at $z\sim0.05$, [Ne{\sc v}] at $z\sim0.15$ and Mg{\sc i} and Mg{\sc ii} at $z\sim0.4$ in our sample. Our photometric redshift selection is conservative as there could be sources with $z_{\rm phot}$ in the $1.2-1.4$ range which could be either \C or \CIV (Fig.~\ref{fig:zhist}).

\citet{Sobral2017} used $BzK$ colour selections to further improve the completeness of their Ly$\alpha$ sample and thus include some sources with lower photometric redshifts. If we did not remove the Ly$\alpha$ selected sources, we would select an extra 9 \C and 16 \CIV sources. We remove sources selected as Ly$\alpha$ by \citet{Sobral2017} to ensure a high purity and obtain conservative, but secure samples. Note however, the highly unusual colours of \CIV emitters (see Section~\ref{sec:colcol}), which means that some real \CIV emitters might have been selected as Ly$\alpha$ and thus were removed from our sample. Spectroscopic follow up is required to further investigate this. See also discussion in \citet{Sobral2017} on the removal of the vast majority of \C and \CIV contaminants in CALYMHA, which is usually not done in other Ly$\alpha$ surveys.

\begin{figure*}
\centering
\includegraphics[trim=0cm 0cm 0cm 0cm, width=0.999\textwidth]{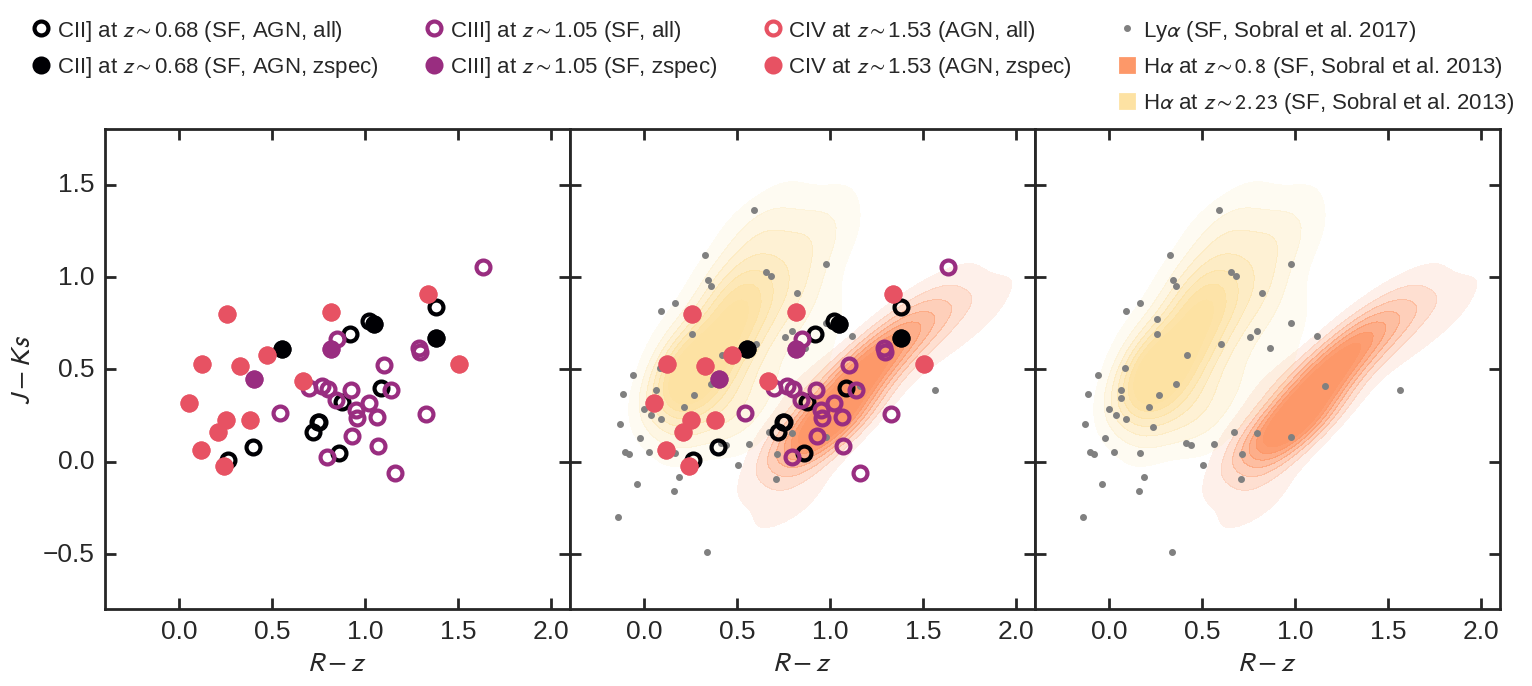}
\caption{$R-z$ versus $J-Ks$ colour-colour plot for the emitters. Ly$\alpha$ emitters at $z\sim2.23$ from  the CALYMHA survey \citep{Sobral2017} and H$\alpha$ emitters from the HiZELS project \citep{2013MNRAS.428.1128S} are also plotted. \CC and \C emitters are located in the $z\sim0.8-1.5$ SF galaxy regime. However, the \CIV emitters have unusual colours, following the distribution of SF at $z\sim2.23$, rather than $1.5$.}
\label{fig:colcol}
\end{figure*}

\section{Properties of the \CC, \C and \CIV samples}\label{sec:properties}

Table~\ref{tab:numbers} lists the final samples of \CC, \C and \CIV emitters, which include spectroscopically confirmed sources and sources selected through their $z_{\rm phot}$.  We have $3$ spectroscopically confirmed \CC emitters in addition to $13$ $z_{\rm phot}$. In the case of \C emitters, we have $4$ spectroscopically selected sources and $30$ with $z_{\rm phot}$. We obtain $14$ \CIV sources with $z_{\rm spec}$ and $3$ with $z_{\rm phot}$. Note the particularly high spectroscopic completeness of the \CIV sample, a likely result of the follow up of Chandra COSMOS sources.

In this section, we study the colour-colour properties as well as the colour and EW distributions with the aim of investigating the nature of the \CC, \C and \CIV emitters, as well as test the robustness of our sample.  We also investigate the properties of the emitters using X-ray, radio and space telescope optical data. Tables~\ref{tab:EW} and~\ref{tab:summary} summarise the EW, UV, optical, X-ray and radio properties of the sample, while Tables~\ref{tab:indivCII}, \ref{tab:indivCIII} and \ref{tab:indivCIV} describe individual \CC, \C and \CIV emitters. We list their sky coordinates, line luminosity, rest-frame $EW_{\rm rest}$, observed $(U-B)_{\rm obs}$ colours and describe their optical {\it HST} morphologies, and their X-ray and/or radio counterparts. Note that while most of the COSMOS part of the CALYMHA survey is covered by {\it HST} and VLA radio data, the deep Chandra data is only available for a sub-area. In the case of UDS, only {\it HST} data is available and for a small subarea of the field.

To describe the properties of our \CC, \C and \CIV sources, we will use the emission line luminosity, which is derived from the observed flux measured within $3''$ apertures \citep[which corresponds to about 30 kpc at the redshift of our emitters, see also][]{Sobral2017}:
\begin{equation}
\label{eq:L}
L_\mathrm{line} = 4 \pi D^2_{L}(line) F_\mathrm{line},
\end{equation}
where $\mathrm{line}$ is \CC, \C or \CIV and $D_{L}(line)$ is the luminosity distance at the redshift of each line (see Table~\ref{tab:lines}). 

\subsection{Colour-colour properties}\label{sec:colcol}

In SF galaxies, the $R-z$ versus the $J-Ks$ colour space probes the $4000$\,{\AA} break, which moves from between the $R$ and $z$ filters for sources at $z\sim0.7-1.2$, to between $J$ and $Ks$ for sources at $z>2.1$. In Fig.~\ref{fig:colcol}, this is illustrated by the population of SF H$\alpha$ emitters from \citet{2013MNRAS.428.1128S} which move from the lower-right side at $z\sim0.8$ (large $R-z$, small $J-Ks$) of the plot towards the upper-right side (small $R-z$, larger $J-Ks$) at $z\sim2.2$. For comparison, we also overplot the CALYMHA Ly$\alpha$ emitters at $z\sim2.2$.

Some \CC emitters at $z\sim0.7$ are located in the colour space of SF galaxies at $z\sim1.5$, so they have atypical colours for their redshift, indicating that while some may trace SF galaxies, some, as expected, probably result from ionisation in AGN through shocks.

\C sources mostly lie in the region of $z\sim0.8-1.0$ SF galaxies, possibly indicating a SF, rather than AGN nature of these emitters. Note that the most extreme \C emitter, with a very low $R-z$ colour, is an AGN (see Section~\ref{sec:ancillary}) and the most luminous in the emission line ($\sim10^{42.4}$ erg\,s$^{-1}$). 

Given the large ionisation energy necessary to produce \CIV, it is expected that \CIV requires either an AGN or very hot stars. It is therefore perhaps not surprising that most \CIV emitters at $z\sim1.5$ do not have colours consistent with SF galaxies at that redshift, but lie in the region of $z\sim2.2$ Ly$\alpha$ emitters. Note that all \CIV sources with all the two colours required for Fig.~\ref{fig:colcol} are spectroscopically confirmed, so the unusual colours cannot be attributed to a wrong selection. It is therefore crucial to consider the contamination by \CIV emitters to samples of NB selected Ly$\alpha$ emitters: without redshifts, when using colours, many lower redshift \CIV sources will be confused with higher redshift Ly$\alpha$ emitters, as noted by \citet{Sobral2017}. \CIV have unusual spectral shapes in other bands as well. For example, the criteria \citet{2016ApJ...823...20K} used for selecting Ly$\alpha$ emitters at $z\sim2.2$ ($(U-NB)>0.5$ and $(B-NB)>0.2$), would select 7 spectroscopically confirmed \CIV emitters as Ly$\alpha$ emitters. Note this corresponds to half the sample of confirmed \CIV emitters. Even when using the criteria defined by \citet{2016ApJ...823...20K} to select `secure' Ly$\alpha$ emitters ($(U-NB)>0.9$ and $(B-NB)>0.2$), we would still select 5 spectroscopically confirmed \CIV into a Ly$\alpha$ sample. These \CIV emitters are typically luminous, so they result in contamination of the bright end of the Ly$\alpha$ distribution \citep{Sobral2017, Matthee2017}. Because this contamination is mostly at bright fluxes, it is less important in deep but small area NB surveys \citep[e.g.][]{2015ApJ...809...89T}.

\renewcommand{\arraystretch}{1.2}
\begin{table*}
\begin{center}
\caption{Rest-frame $EW_{\rm rest}$, observed $(U-B)_{\rm obs}$ and UV slope $\beta$ of the emitters which have photometric or spectroscopic redshifts. We also list the observed filters used for tracing the rest-frame UV. Note that the rest-frame wavelength traced for calculating the $\beta$ slopes vary slightly depending on the emitter type. See Section~\ref{sec:beta} for more details. The uncertainties reported represent the standard deviation of the sample.}
\begin{tabular}{l c c c c c c c c c}
\hline\hline
Line & $z_{\rm spec}$ & Mean $EW_{\rm rest}$ & Median $EW_{\rm rest}$ & Mean $(U-B)_{\rm obs}$ & Median $(U-B)_{\rm obs}$ & Mean $\beta$ & Median $\beta$ & Filters for \\
& & (\AA) & (\AA) & (mag) & (mag) & & & $\beta$ slope  \\
\hline
\CC & $0.68$ & $82\pm56$ & $74\pm70$ & $0.25\pm0.20$ & $0.26\pm0.25$ & $-2.0\pm0.4$ & $-1.9\pm0.5$ & $NUV$, $U$ \\
\C  & $1.05$ & $93\pm59$ & $87\pm74$ & $0.25\pm0.37$ & $0.21\pm0.46$ & $-0.8\pm1.1$ & $-0.6\pm1.4$ & $U$, $B$ \\
\CIV & $1.53$ & $51\pm46$ & $34\pm58$ & $0.15\pm0.37$ & $0.21\pm0.46$ & $-1.9\pm0.8$ & $-1.6\pm1.0$ & $U$, $V$ \\ \hline
Ly$\alpha$ & $2.23$ & $85\pm57$ & $77\pm71$ & $0.18\pm0.25$ & $0.23\pm0.31$  & $-1.6\pm0.6$ & $-1.7\pm0.7$ & $g$, $R$\\
H$\alpha$ & $2.23$ &            &      &               &       & $-1.0\pm0.6$ & $-1.0\pm0.7$ & $g$, $R$\\
\hline
\end{tabular}
\label{tab:EW}
\end{center}
\end{table*}
\renewcommand{\arraystretch}{1.1}

\subsection{Observed and rest-frame $U-B$ colours}\label{sec:UBcolor}

\begin{figure}
\centering
\includegraphics[trim=0cm 0cm 0cm 0cm, width=0.479\textwidth]{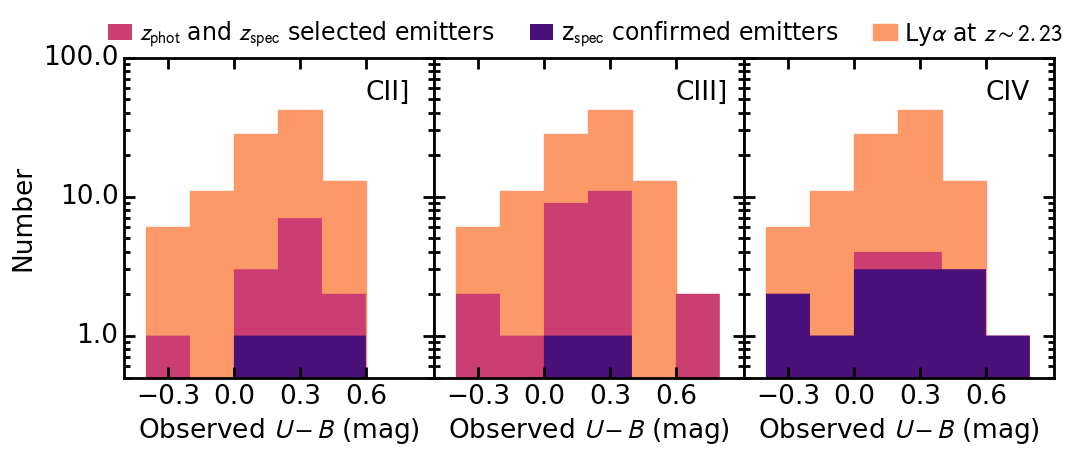}
\caption{Distribution of observed $(U-B)_{\rm obs}$ colours of the emitters. The distribution of the Ly$\alpha$ emitters at $z\sim2.23$ \citep{Sobral2017} is given for reference. The $(U-B)_{\rm obs}$ average for the C emitters at $0.7 \lesssim z\lesssim 1.5$ indicate blue colours, consistent with Ly$\alpha$ emitters at $z\sim2.23$. }
\label{fig:UB}
\end{figure}

Fig.~\ref{fig:UB} displays the distribution of observed $(U-B)_{\rm obs}$ colours for our emitters, while the mean and median colours are listed in Table~\ref{tab:EW}. For individual sources, the numbers are listed in Tables~\ref{tab:indivCII}, \ref{tab:indivCIII} and \ref{tab:indivCIV}. At $z\sim0.7-1.5$, $(U-B)_{\rm obs}$ approximately traces the rest-frame UV. Note that all but 2 \C emitters have both $U$ and $B$ measurements. Note however that $U$ is contaminated by the emission line, thus the colours should be interpreted with caution. The $(U-B)_{\rm obs}$ corresponds to approximately:
\begin{itemize}
\item \CC: restframe 1380\,{\AA} and 2275\,{\AA}
\item \C: restframe 1865\,{\AA} and 2175\,{\AA}
\item \CIV: restframe 1510\,{\AA} and 2165\,{\AA}
\end{itemize}

All three types of emitters studied in this paper have relatively blue colours with mean $(U-B)_{\rm obs}$ colours in the $0.15-0.25$ range. Our emitters are consistent in colour to Ly$\alpha$ emitters selected at $z\sim2.23$ in the CALYMHA survey, which stand at a mean of $0.18$. For the \C and \CC source with disky morphology as discussed in the previous section, the blue UV colours indicate a relatively dust-free environment. 

\begin{figure}
\centering
\includegraphics[trim=0cm 0cm 0cm 0cm, width=0.299\textwidth]{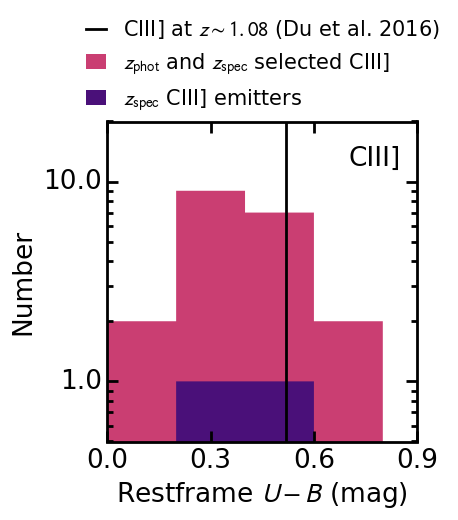}
\caption{Distribution of rest-frame $(U-B)_{\rm rest}$ colours (observed $i-z$ colours) of the \C emitters. The rest-frame $(U-B)_{\rm obs}$ from \citet{2017ApJ...838...63D} averages are only slightly larger, but consistent with our results. This shows that our \C emitters are a relatively blue population of sources, but that our blind selection may also be recovering a few sources which are even bluer than the average.}
\label{fig:iz}
\end{figure}

For comparison, we also present a rest-frame $(U-B)_{\rm rest}$ colour, derived from the observed $i-z$ colour (Fig.~\ref{fig:iz}). Our \C emitters are on average slightly bluer ($(U-B)_{\rm rest}=0.38$) than the spectroscopic sample of \C emitters at $z\sim1$ from \citet{2017ApJ...838...63D}, which characterise their emitters as being blue and low-mass with little dust extinction. Note however that the two distributions are perfectly compatible within the full distribution of values.

\subsection{UV slope $\beta$}\label{sec:beta}
The dust in a galaxy absorbs the UV radiation coming from an AGN or from massive, young stars. Despite it depending on many other properties \citep[see for example][]{2009ApJ...705..936B}, the slope of the rest-frame UV continuum ($\beta$) is usually used as a simple tracer of the dust extinction in a galaxy. Here, we estimate $\beta$ for our \CC, \C and \CIV emitters, defined in the following way:

\begin{equation}
\label{eq:beta}
\beta = - \frac{m_1-m_2}{2.5 \log_{10}\left(\frac{\lambda_1}{\lambda_2}\right)} - 2,
\end{equation}
where $m_1$ and $m_2$ are the magnitudes of the source in two observed filters which trace the rest-frame UV, preferentially around the $1500$\,{\AA} reference wavelength. $\lambda_1$ and $\lambda_2$ are the central wavelengths of the two filters.

Given that our emitters are at 3 different redshifts, it is not simple to have filters that trace exactly the same rest-frame wavelengths. The best choices of filters to trace the rest-frame UV included the $U$ filter, which can be contaminated by the emission lines. We chose filters to match the convention used in other studies and ease comparisons: 
\begin{itemize}
\item \CC: NUV, U
\item \C: U, B
\item \CIV: U, V
\end{itemize}

Note that for \C emitters, the $U$ filter traces a slightly redder rest-frame wavelength compared to other studies and \CC and \CIV, which may bias $\beta$ to slightly redder values.

We list the averages of the $\beta$ slope in Table~\ref{tab:EW}. The $\beta$ slopes of our emitters indicate a steep UV continuum with potential low dust attenuation, within the same ranges as Ly$\alpha$ emitters at $z\sim2.23$, but with \C being slightly redder.

\begin{figure}
\centering
\includegraphics[trim=0cm 0cm 0cm 0cm, width=0.479\textwidth]{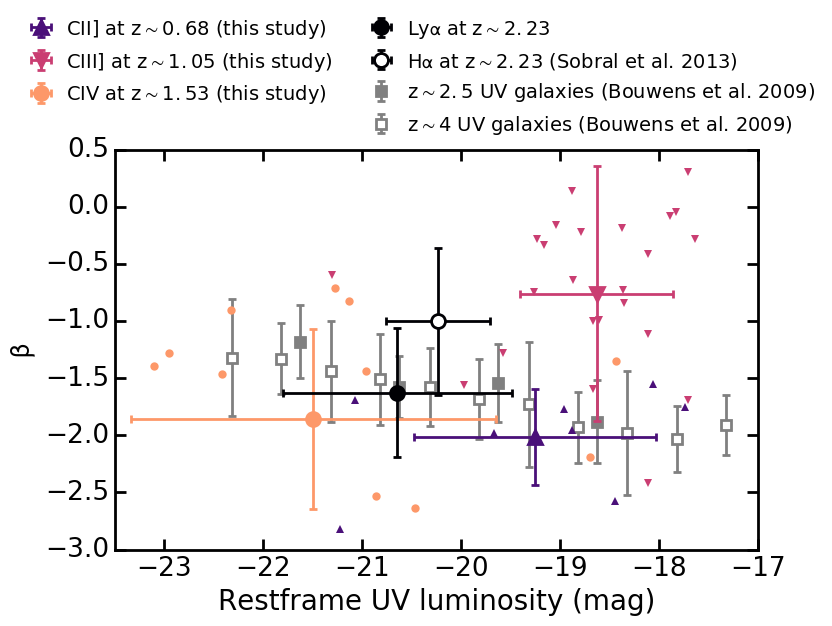}
\caption{Distribution of rest-frame UV slope $\beta$ as function of UV magnitude for our sample of emitters and Ly$\alpha$ \citep{Sobral2017} and H$\alpha$ at $z\sim2.23$ \citep{2013MNRAS.428.1128S}. Only sources with both rest-frame UV bands detected are plotted. Values for individual \CC, \C and \CIV emitters are plotted in the smaller symbols, while averages are shown in the larger symbols with error bars. For comparison, we plot the values for $z\sim2.5$ and $z\sim4$ Lyman break, UV selected galaxies from \citet{2009ApJ...705..936B}. \CC and \CIV emitters have colours consistent with the population of UV selected galaxies. \C emitters are redder, consistent with more general populations of SF galaxies such as those selected through H$\alpha$.}
\label{fig:beta}
\end{figure}

We show the average $\beta$ slope compared to the average absolute UV magnitude $M_{\rm UV}$ in Fig.~\ref{fig:beta}. The relation between the UV slope and absolute UV magnitude has been shown to not depend significantly on redshift for LBGs, thus making it a good probe for studying galaxies at all cosmic epochs \citep{2009ApJ...705..936B, 2012ApJ...756...14S}. 

Fig.~\ref{fig:beta} shows that the \CC and and \CIV emitters are consistent in UV properties with other Lyman break selected SF galaxies at higher redshift \citep[$z\sim2.5-4$,][]{2009ApJ...705..936B} and with Ly$\alpha$ selected galaxies at $z\sim2.2$. It is important to note that Ly$\alpha$ and LBG selected samples are generally biased towards blue, less massive, metal-poor SF galaxies \citep{2015MNRAS.452.2018O}. Note that Ly$\alpha$ can also probe extremely dusty galaxies unlike the LBG technique \citep{2015MNRAS.452.2018O, 2016MNRAS.458..449M}. The average $\beta$ slope is consistent with the results obtained from the observed and intrinsic colours, indicating that \CC and \CIV emitters are relatively blue objects with little dust extinction. At first glance, this is quite surprising, because, as was discussed in previous sections, a large fraction of \CC emitters and the bulk of \CIV emitters have properties consistent with AGN. However, young, dust-free, quasar-like AGN will have steep UV continua, similar to those measured for \CIV and \CC emitters. 

\C emitters have relatively flat slopes, indicating a redder UV continuum. This means \C emitters have $\beta$ slopes consistent with more general populations of SF galaxies such as those selected from H$\alpha$ \citep{2015MNRAS.452.2018O}. The \C emission line may therefore be a good, unbiased tracer of SF galaxies with a range of properties.

\subsection{EW$_\mathrm{rest}$ distribution}\label{sec:EW}

We also investigate the distribution of rest-frame $EW_\mathrm{rest}$ in the sample of emitters. We find that the average EWs are high. This could be caused by Ly$\alpha$ interlopers, which can have large observed EWs. 

Therefore, as a further conservative step, we attempt to bring any contamination from Ly$\alpha$ emitters (which may have high EWs) to virtually zero. We do this by applying colour cuts targeting the Lyman break in $z\gtrsim2$ galaxies to further remove any potential Ly$\alpha$ interlopers from the sample:
\begin{equation}
(NUV - U) > 1.0 \quad \mathrm{or} \quad (NUV - B) > 1.5
\end{equation}

\begin{figure}
\centering
\includegraphics[trim=0cm 0cm 0cm 0cm, width=0.479\textwidth]{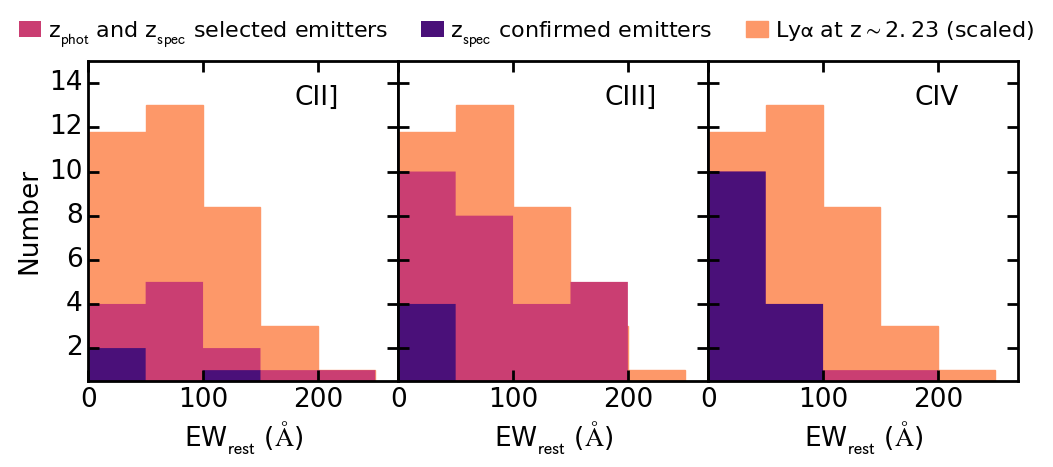}
\caption{Rest-frame EW distribution for highly secure \CC, \C, \CIV, classified as such by spectroscopic or photometric redshifts. We removed any potential high redshift sources which were classified as C species based on $z_{\rm phot}$, but as $z>2$ sources by colour-colour selections from \citet{Sobral2017} or by Lyman break colour cuts. Note the large average EW for all three emitter types. For comparison, we also show the EW distribution of the Ly$\alpha$ emitters at $z\sim2.23$ selected in \citet{Sobral2017} (scaled by 0.2).}
\label{fig:EW}
\end{figure}

The distributions of the resulting samples with very high purity are given in Fig.~\ref{fig:EW}. Averages are listed in Table~\ref{tab:EW}, while individual values are given in Tables~\ref{tab:indivCII}, \ref{tab:indivCIII} and \ref{tab:indivCIV}. 

We find a significant population with large rest-frame EWs for all three emitter species, potentially extending up to $200$\,{\AA}. The distributions of the three populations drop in numbers towards high EWs. The small bump in the distribution of \C emitters in the largest EW bin is not statistically significant. On average, \CIV emitters have lower EWs compared to \CC and \C, which correlates with the brighter average magnitudes of these \CIV sources (see for example, Fig.~\ref{fig:beta}). This is caused by the prevalence of quasars among \CIV sources. Overall, there are relatively few sources with very low EWs close to our selection limit of $16$\,{\AA}. Because of the way the selection of emitters is performed \citep[see for details][]{Sobral2017}, sources will be classified as emitters if they pass the EW limit cut of $16$\,{\AA} and a signal-to-noise cut. For low S/N sources, the EW needs to be higher for a source to pass the selection compared to a source which is bright in the BB. This explains the proportionally larger number of \CIV emitters with low EW compared to \CC and \C sources. We would like to note that while EWs can be biased high because of a number of geometrical reasons (see discussion below, e.g. offset of emitting line regions with respect to underlying UV radiation), the line fluxes given our apertures will be reliable, as all the emission should be captured, more so when compared to, for example slit observations.

Another explanation for the larger EWs would be that EWs measured from the NB are not accurate with respect to EWs from spectroscopy. However, we do not believe this could be the case as NB EWs have been found to be reliable when compared to spectroscopic observations \citep[e.g. for lines such as H$\alpha$, Ly$\alpha$][]{2015ApJ...808..139S, 2015MNRAS.450..630S,2016MNRAS.458.3443S}, specifically for CALYMHA survey follow-up of Ly$\alpha$ emitters (Sobral et al. in prep). Some \CC and \CIV sources, including those with large EWs, are confirmed spectroscopically, however, none of the large EW \C emitters has a $z_{\rm spec}$. The available spectroscopy is biased towards continuum bright sources, which do not have large EW. Despite potentially strong emission lines with large EW, many continuum-faint \CC, \C and \CIV emitters in our sample, with NB magnitudes below $22-23$, were never spectroscopically followed up precisely because they were not bright enough. While we removed sources of systematic errors and potential interlopers to the best of our abilities, without targeted spectroscopy it is not possible to fully confirm the large EW measurements.

\begin{figure*}
\centering
\includegraphics[trim=0cm 0cm 0cm 0cm, width=0.192\textwidth]{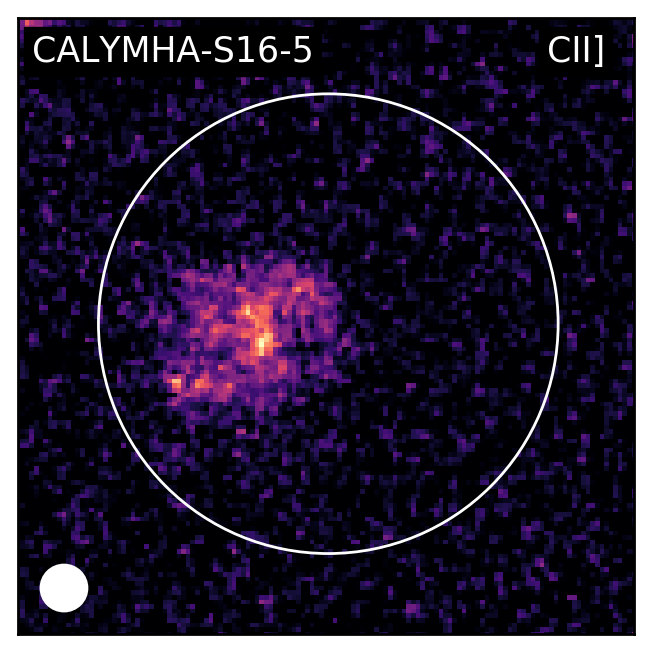}\includegraphics[trim=0cm 0cm 0cm 0cm, width=0.192\textwidth]{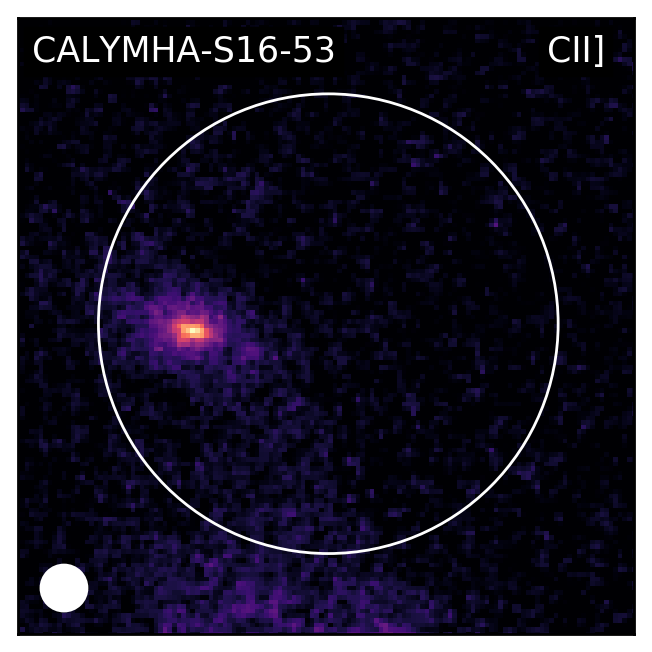}
\includegraphics[trim=0cm 0cm 0cm 0cm, width=0.192\textwidth]{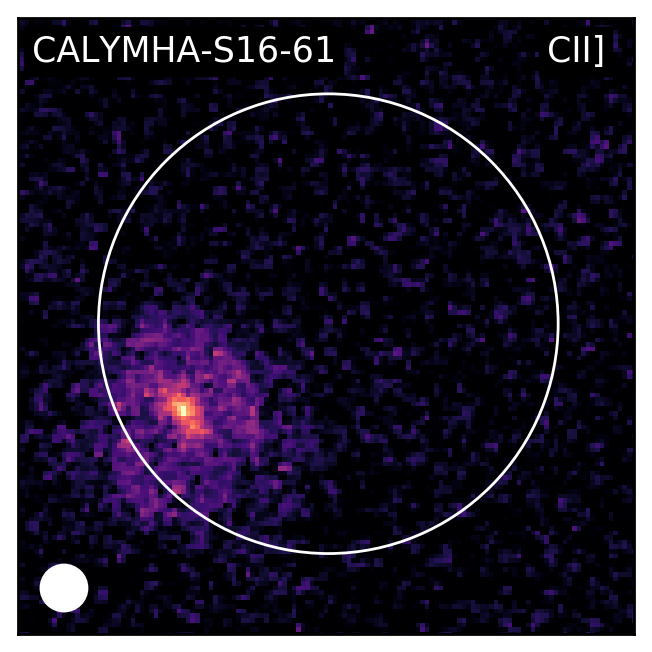}
\includegraphics[trim=0cm 0cm 0cm 0cm, width=0.192\textwidth]{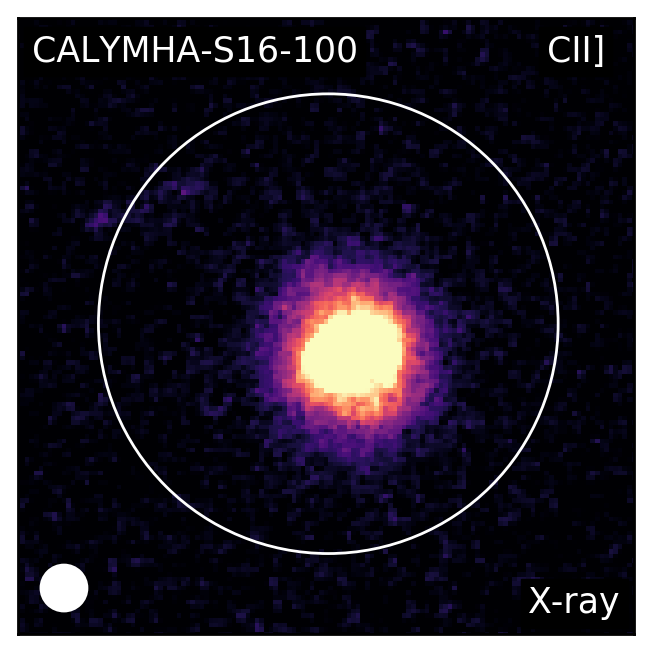}
\includegraphics[trim=0cm 0cm 0cm 0cm, width=0.192\textwidth]{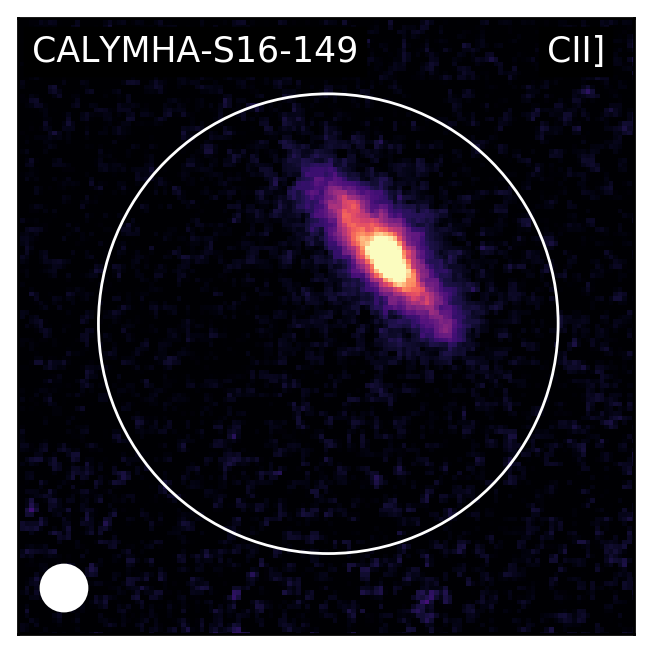}
\includegraphics[trim=0cm 0cm 0cm 0cm, width=0.192\textwidth]{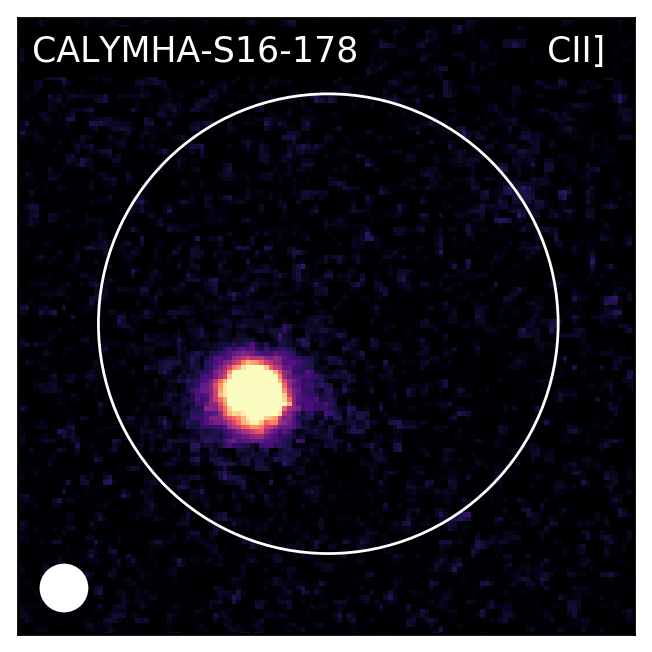}
\includegraphics[trim=0cm 0cm 0cm 0cm, width=0.192\textwidth]{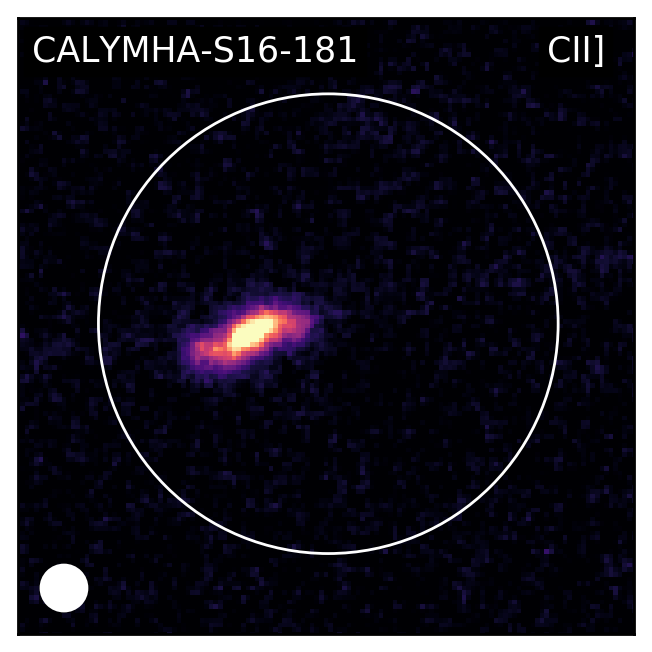}
\includegraphics[trim=0cm 0cm 0cm 0cm, width=0.192\textwidth]{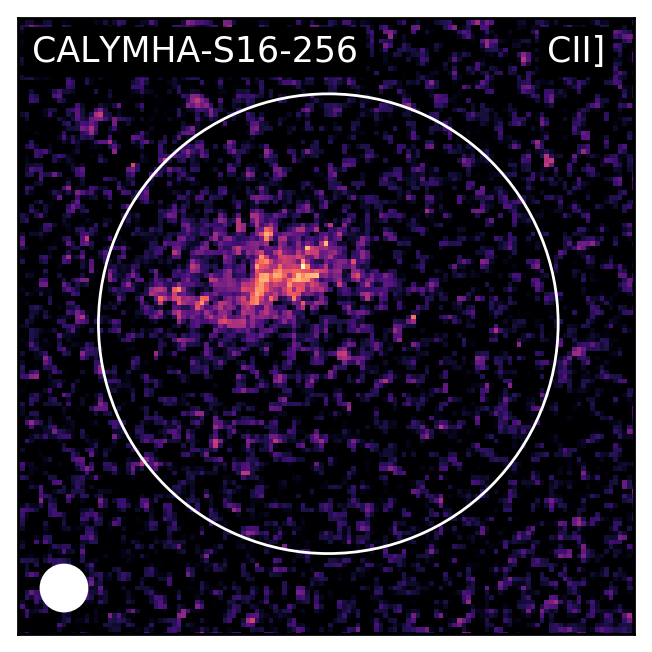}
\includegraphics[trim=0cm 0cm 0cm 0cm, width=0.192\textwidth]{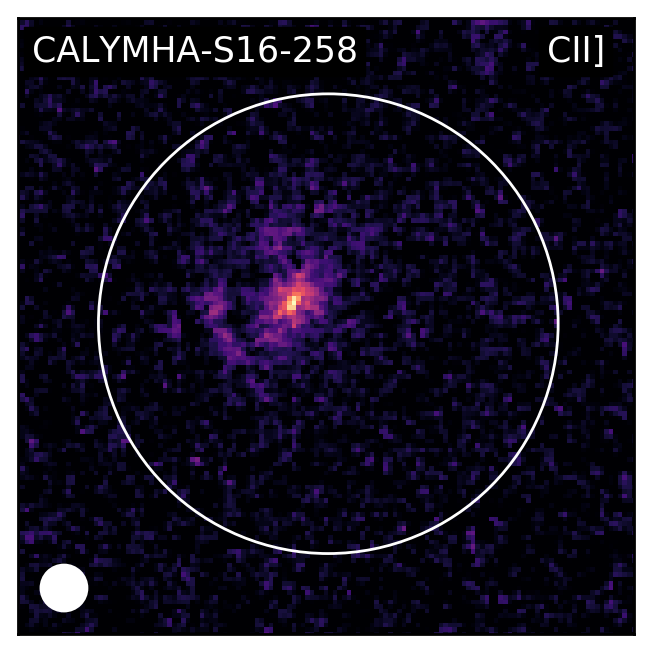}
\includegraphics[trim=0cm 0cm 0cm 0cm, width=0.192\textwidth]{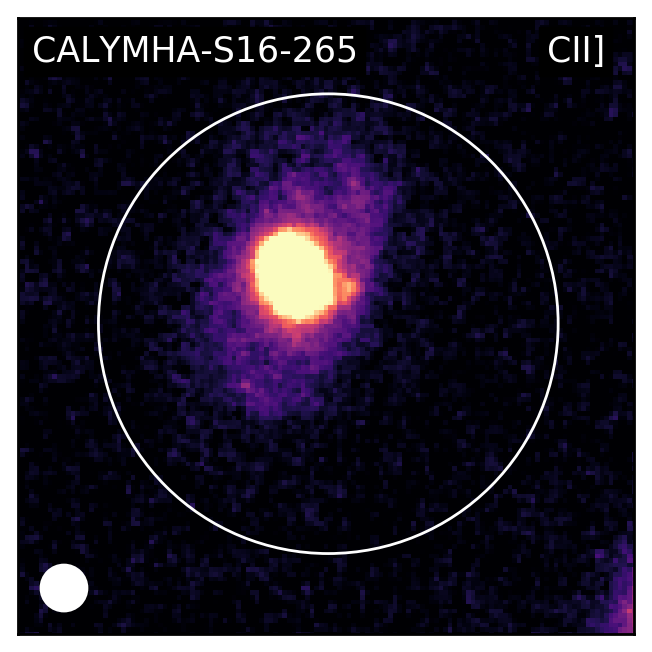}
\includegraphics[trim=0cm 0cm 0cm 0cm, width=0.192\textwidth]{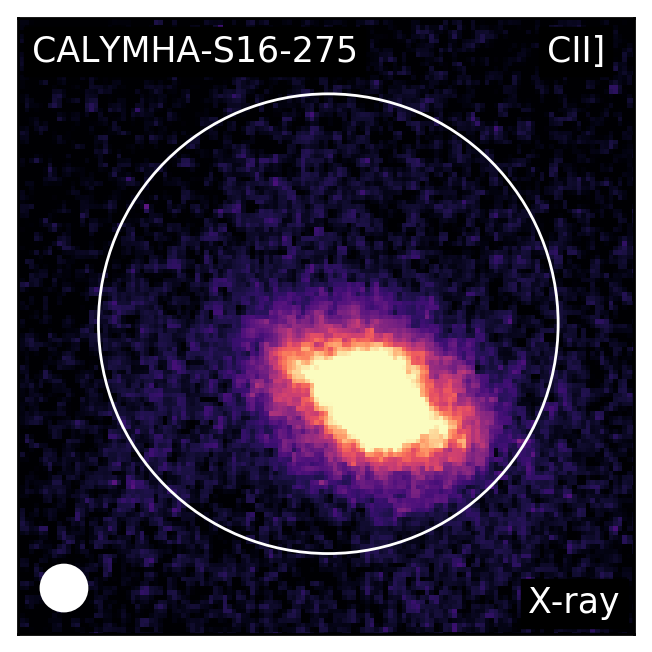}
\includegraphics[trim=0cm 0cm 0cm 0cm, width=0.192\textwidth]{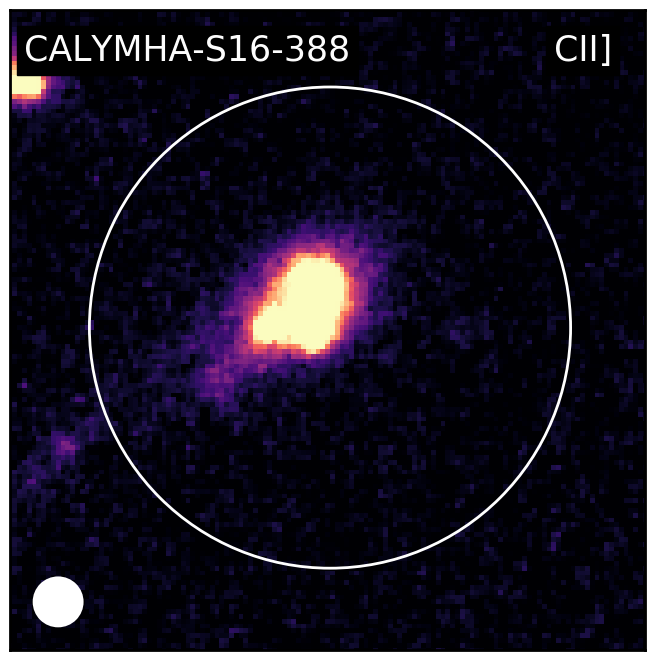}
\caption{{\it HST} cutouts of $z\sim0.7$ \CC emitters selected based on $z_{\rm spec}$ or $z_{\rm phot}$. Images are on the same colour scale, from $0$ to $20\times\sigma_{\rm RMS}$, with a size of $4''\times4''$. At the redshift of \CC emitters, the images have a size of $28.3$\,kpc on each side. The large circle represents the $3''$ aperture used to extract photometry for the CALYMHA sources. The small circle represents $3$ times the {\it HST} point-spread function (PSF $\sim1''$), encompassing 98 per cent of the flux. We also indicate whether the source has an X-ray or radio counterpart. Note that some sources simply do not have X-ray or radio counterparts, while others are not covered with such data. While some \CC emitters have disky morphologies, some have very bright nuclei, which in the case of source 275 correlates with an X-ray detection, indicating these galaxies are Seyfert-like. We center the {\it HST} thumbnails on the peak position of the emission line and find that, in some cases, there are offsets of $\sim5$ kpc from the peak UV rest-frame emission. This may explain the relatively large EW we measure. }
\label{fig:HSTCII}
\end{figure*}

A word of warning is that the errors on the rest-frame EW can be large, with an average of $\sim60$ per cent of the EW value. Hence, some of the values can be $60$ per cent lower or higher than estimated here. Furthermore, as we will discuss in Section~\ref{sec:morphologies}, our morphological results show that we may be tracing specific regions within galaxies with little to no UV continuum, which may bias the EW towards higher values.

Spectroscopic observations are necessary to pose tighter constraints on the EW values, reduce the error bars and further investigate the validity of NB observations for measuring the EW of \CC, \C and \CIV emitters. EW could also be overestimated due to variable sources, which we discuss in detail in Section~\ref{sec:variability}. While variability could explain part of the population, the entire population of large EW sources cannot be explained this way.

\begin{figure*}
\centering
\includegraphics[trim=0cm 0cm 0cm 0cm, width=0.192\textwidth]{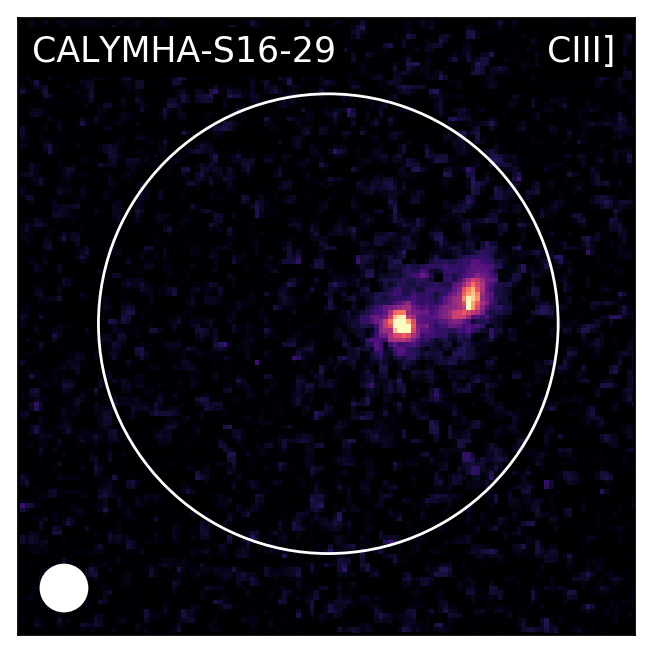}
\includegraphics[trim=0cm 0cm 0cm 0cm, width=0.192\textwidth]{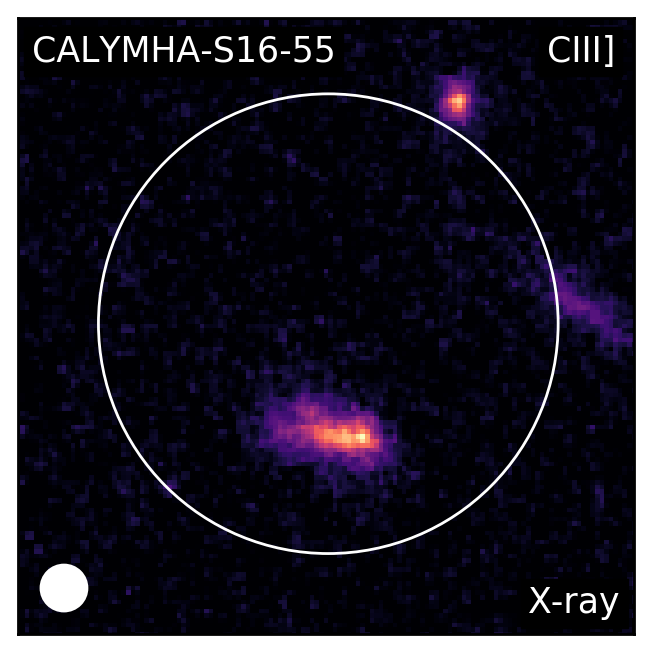}
\includegraphics[trim=0cm 0cm 0cm 0cm, width=0.192\textwidth]{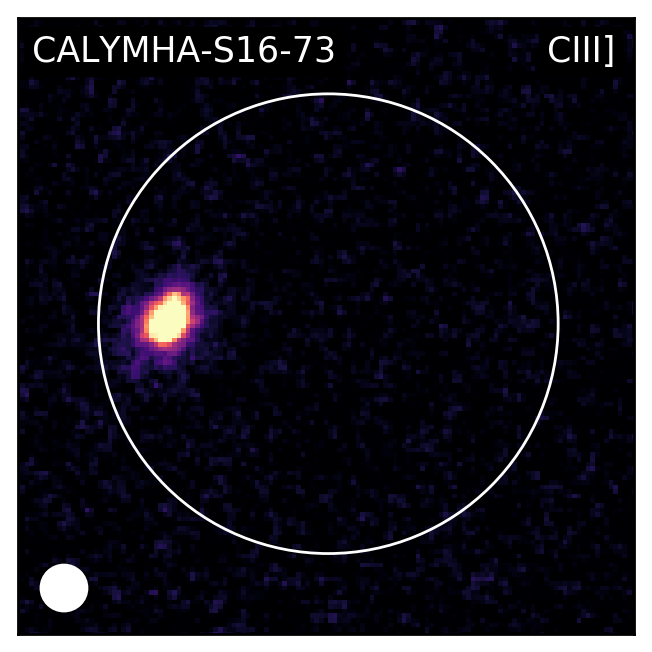}
\includegraphics[trim=0cm 0cm 0cm 0cm, width=0.192\textwidth]{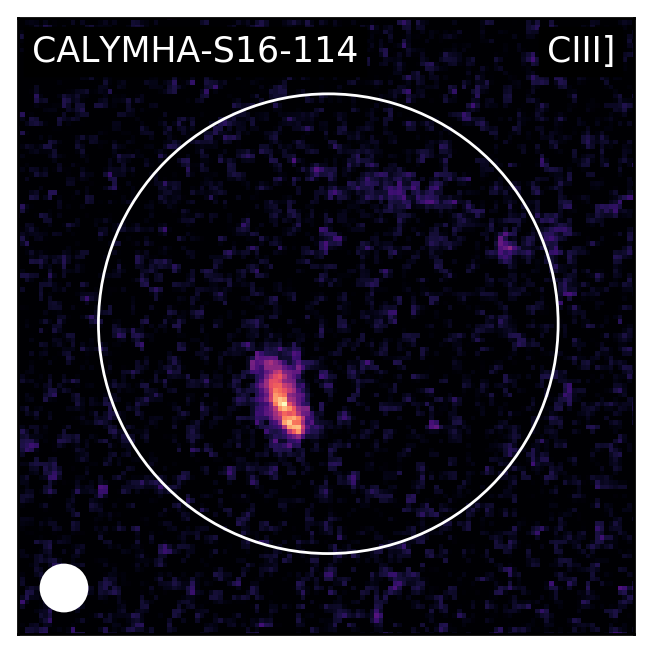}
\includegraphics[trim=0cm 0cm 0cm 0cm, width=0.192\textwidth]{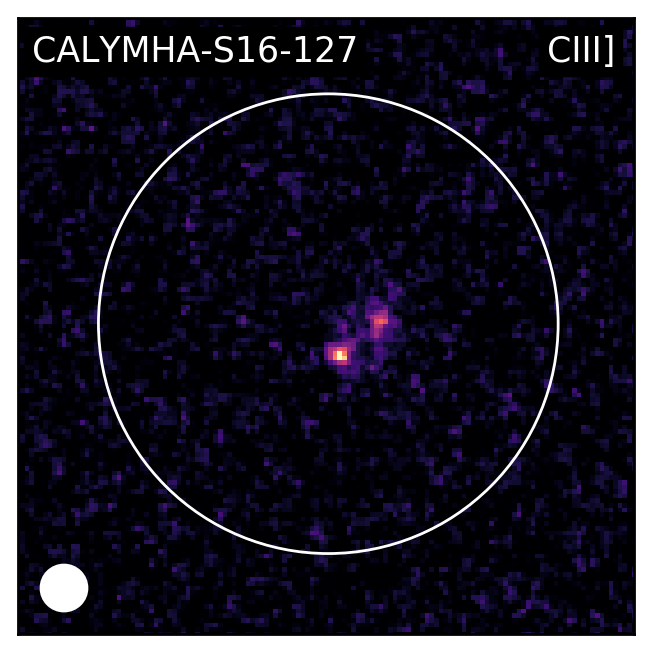}
\includegraphics[trim=0cm 0cm 0cm 0cm, width=0.192\textwidth]{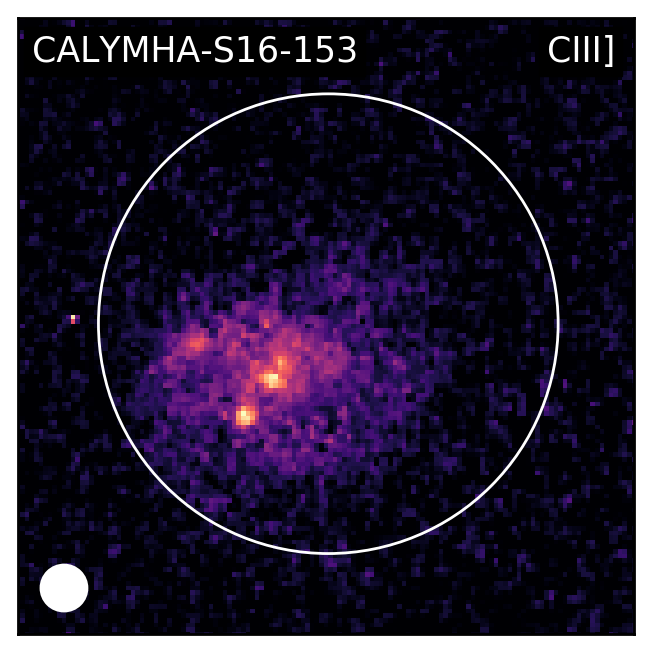}
\includegraphics[trim=0cm 0cm 0cm 0cm, width=0.192\textwidth]{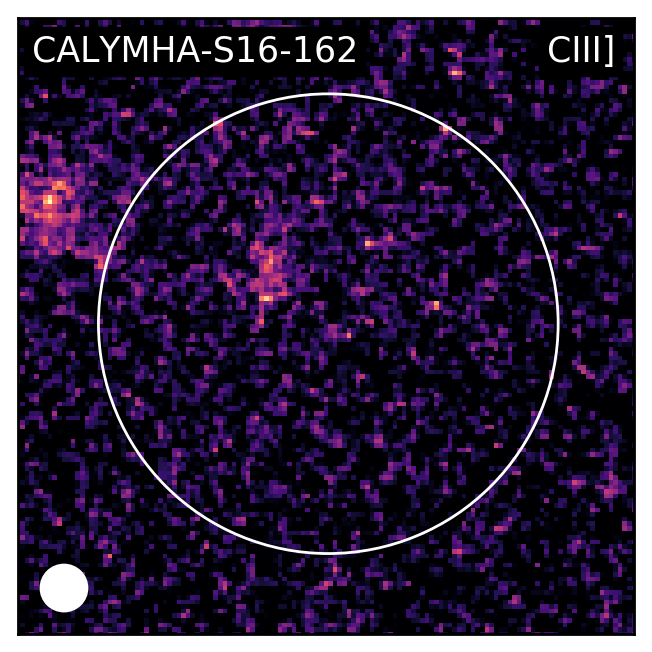}
\includegraphics[trim=0cm 0cm 0cm 0cm, width=0.192\textwidth]{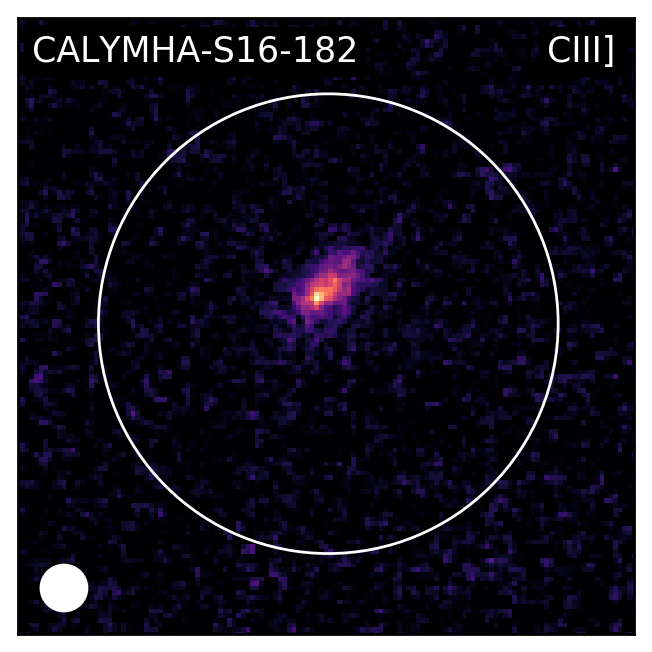}
\includegraphics[trim=0cm 0cm 0cm 0cm, width=0.192\textwidth]{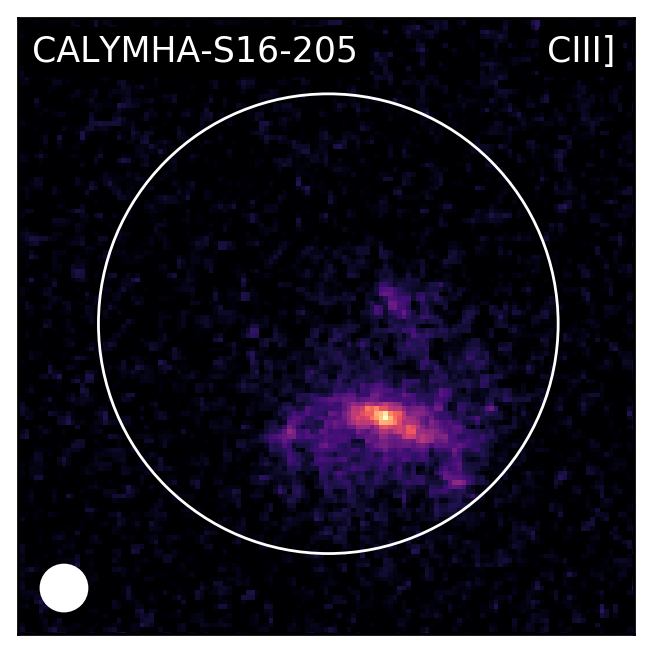}
\includegraphics[trim=0cm 0cm 0cm 0cm, width=0.192\textwidth]{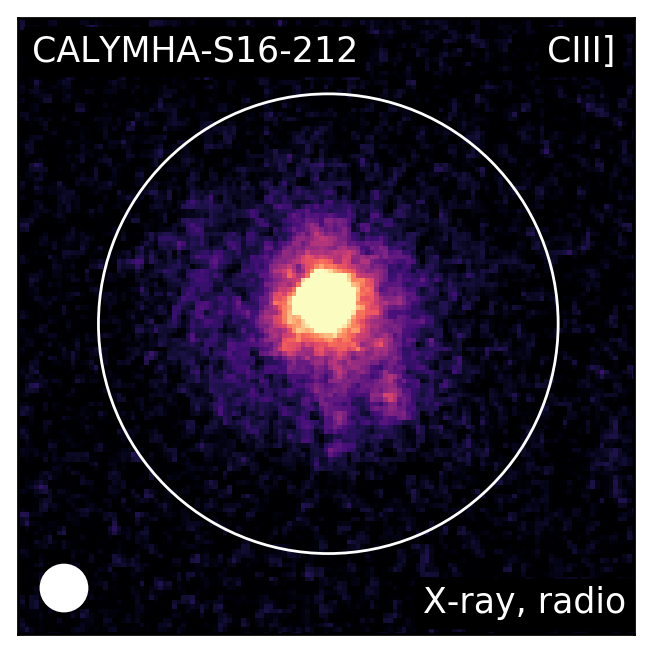}
\includegraphics[trim=0cm 0cm 0cm 0cm, width=0.192\textwidth]{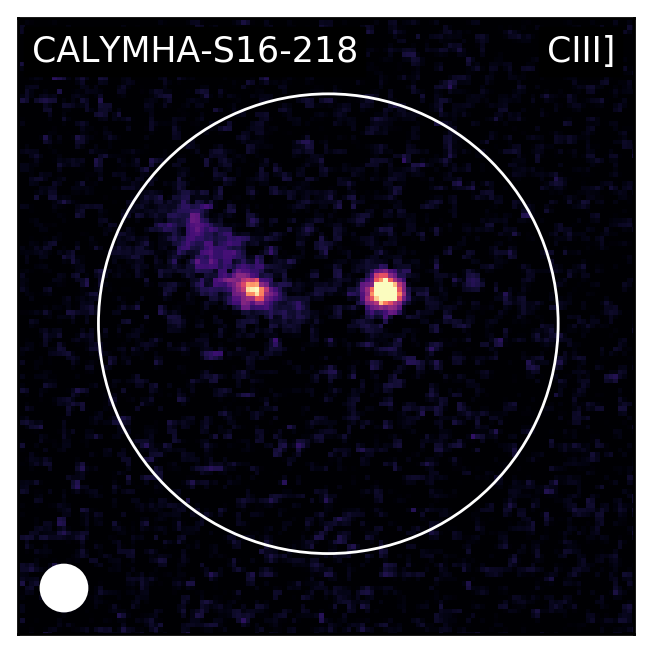}
\includegraphics[trim=0cm 0cm 0cm 0cm, width=0.192\textwidth]{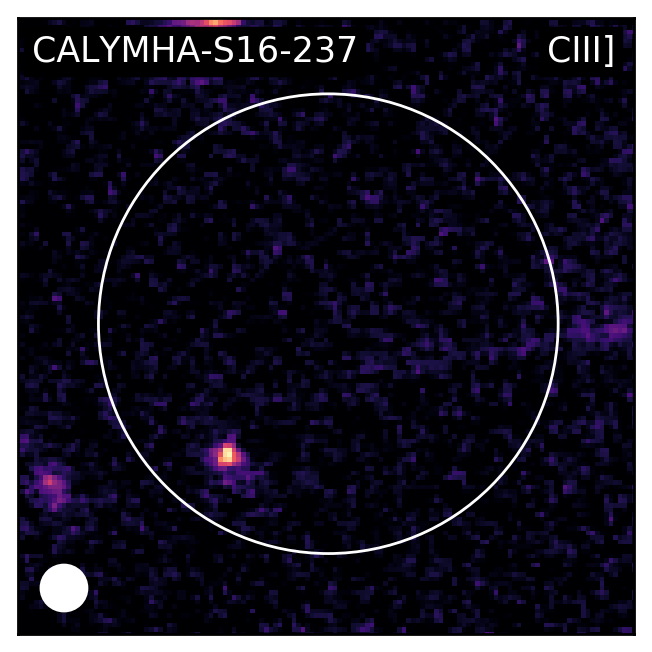}
\includegraphics[trim=0cm 0cm 0cm 0cm, width=0.192\textwidth]{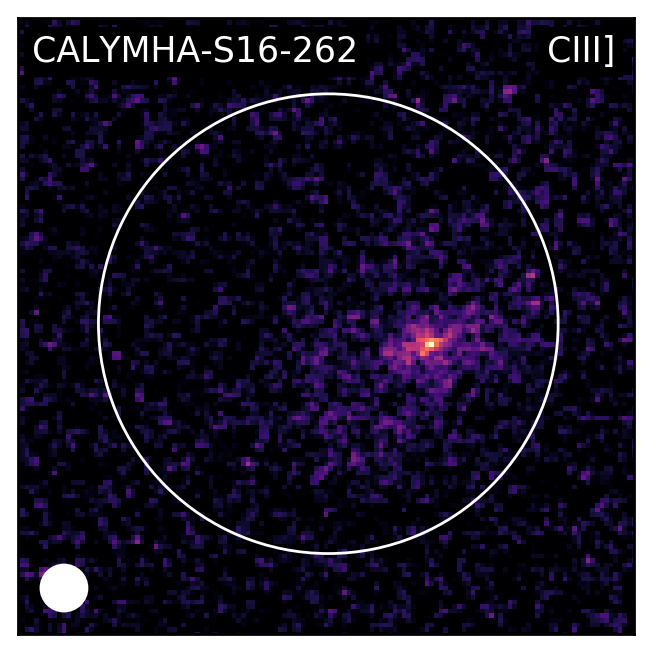}
\includegraphics[trim=0cm 0cm 0cm 0cm, width=0.192\textwidth]{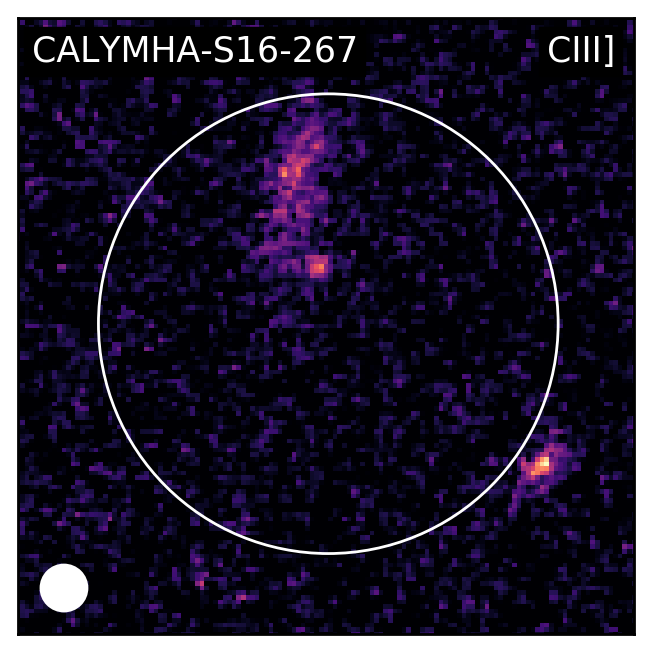}
\includegraphics[trim=0cm 0cm 0cm 0cm, width=0.192\textwidth]{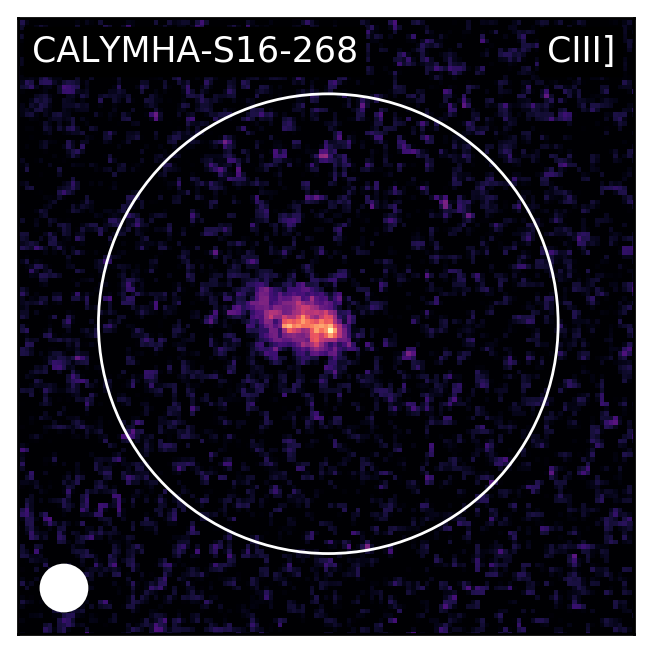}
\includegraphics[trim=0cm 0cm 0cm 0cm, width=0.192\textwidth]{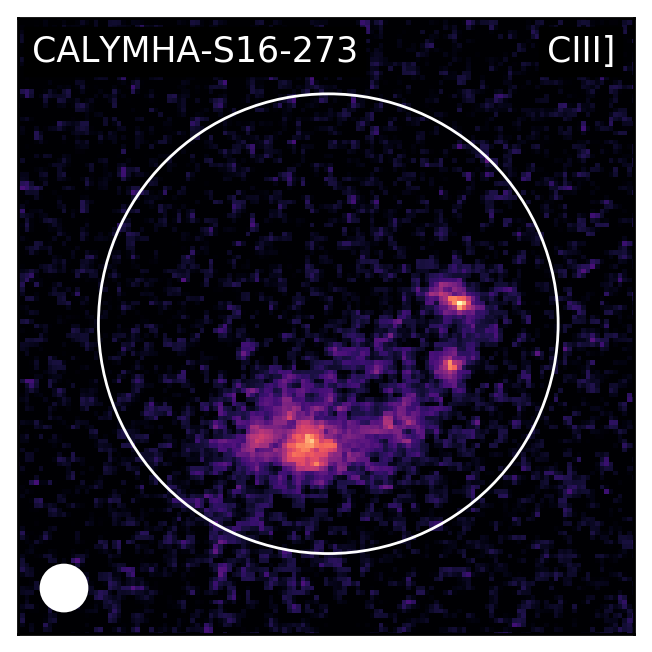}
\includegraphics[trim=0cm 0cm 0cm 0cm, width=0.192\textwidth]{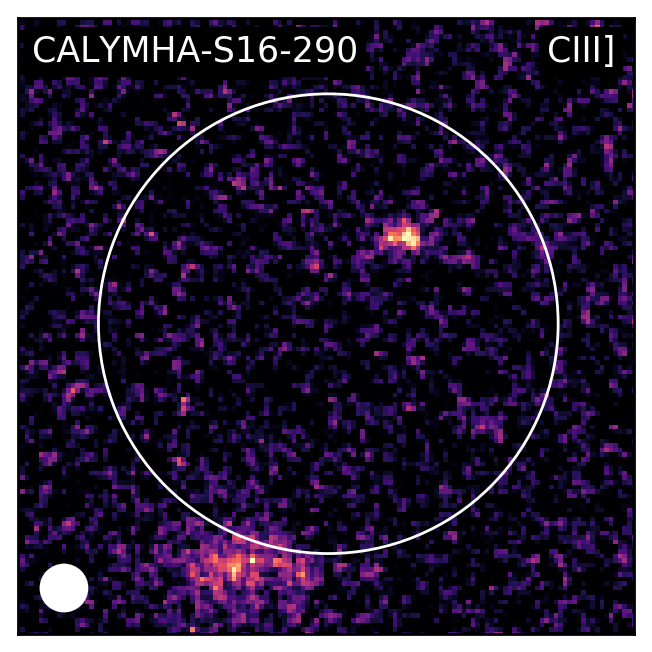}
\includegraphics[trim=0cm 0cm 0cm 0cm, width=0.192\textwidth]{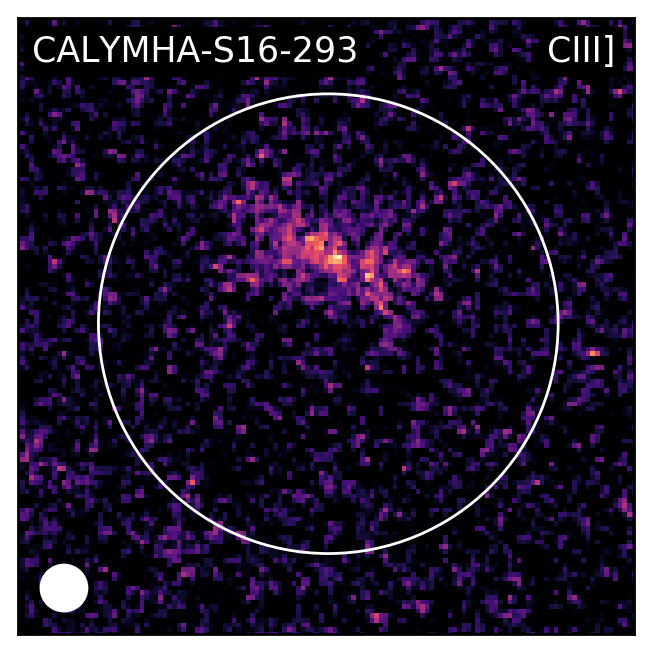}
\includegraphics[trim=0cm 0cm 0cm 0cm, width=0.192\textwidth]{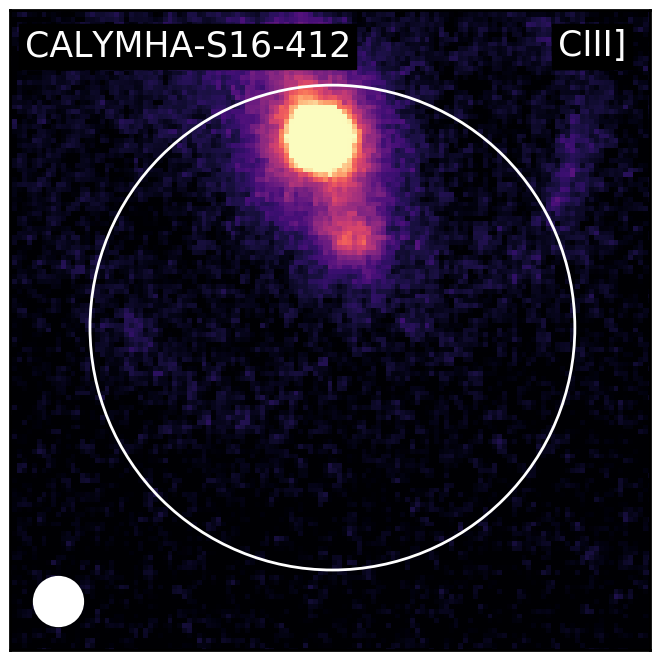}
\caption{Same as Fig.~\ref{fig:HSTCII}, but for \C emitters. The images are $32.4$\,kpc in size at the redshift $z\sim1.05$ of the \C emitters. The large circle is the $3''$ aperture used to extract photometry, while the small circle is $3\times$PSF of {\it HST} point-spread function, which captures 98 per cent of the flux for a point source. \C emitters have a range of optical morphologies, including disks, many interacting or merging sources and messy, complicated, disturbed galaxies. Source 212 has a Seyfert-like morphology with a bright optical core and a radio and X-ray counterpart. Note that in many cases the rest-frame continuum is offset from the CALYMHA emission line detection, which may explain the large $EW$ we find. This is likely caused by physical offsets between the stellar continuum and the brightest emission line regions. A small fraction of sources with coverage have either X-ray or radio detections, indicative of an AGN.}
\label{fig:HSTCIII}
\end{figure*}

\begin{figure*}
\centering
\includegraphics[trim=0cm 0cm 0cm 0cm, width=0.192\textwidth]{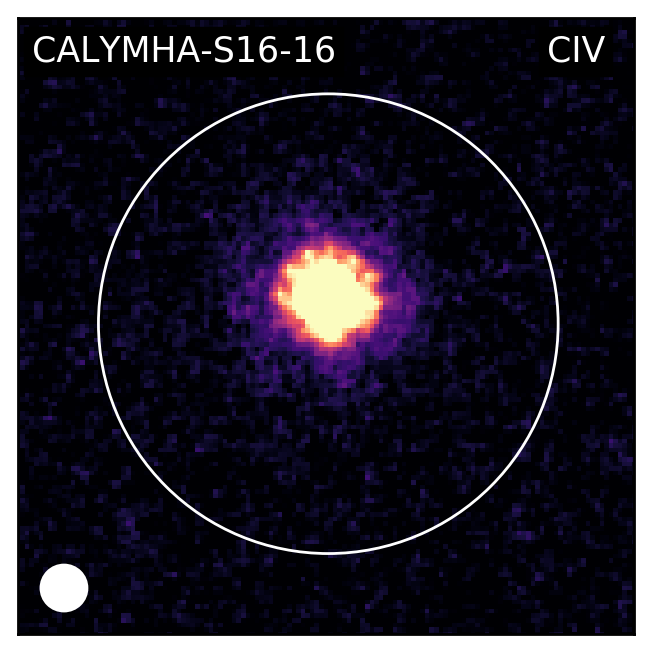}
\includegraphics[trim=0cm 0cm 0cm 0cm, width=0.192\textwidth]{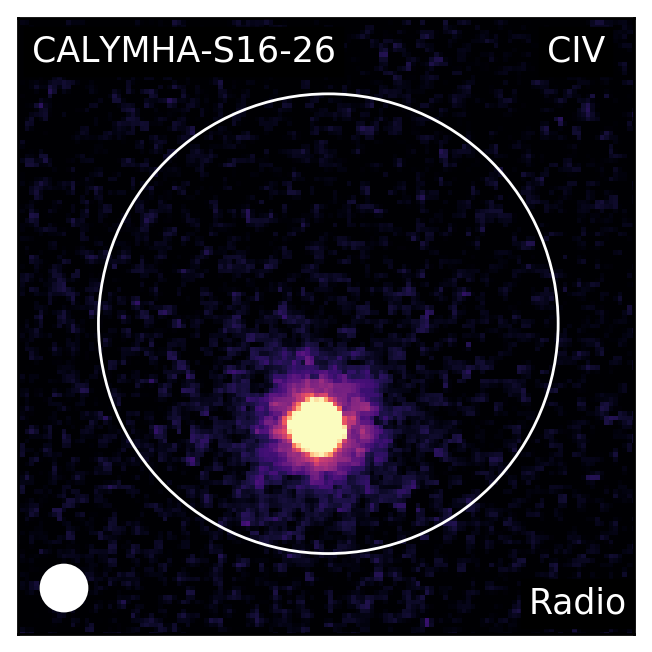}
\includegraphics[trim=0cm 0cm 0cm 0cm, width=0.192\textwidth]{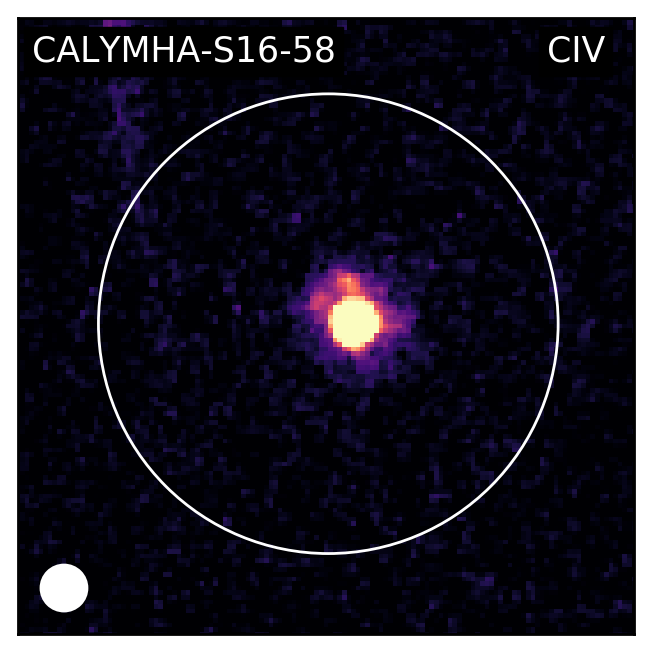}
\includegraphics[trim=0cm 0cm 0cm 0cm, width=0.192\textwidth]{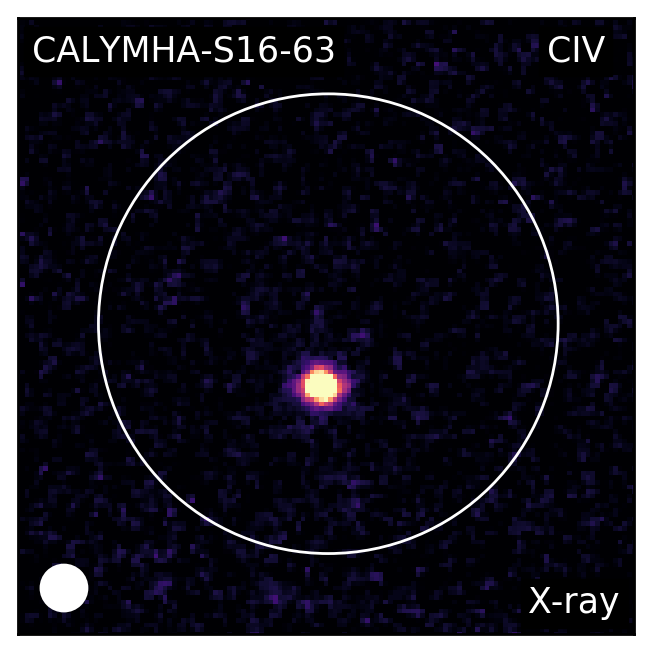}
\includegraphics[trim=0cm 0cm 0cm 0cm, width=0.192\textwidth]{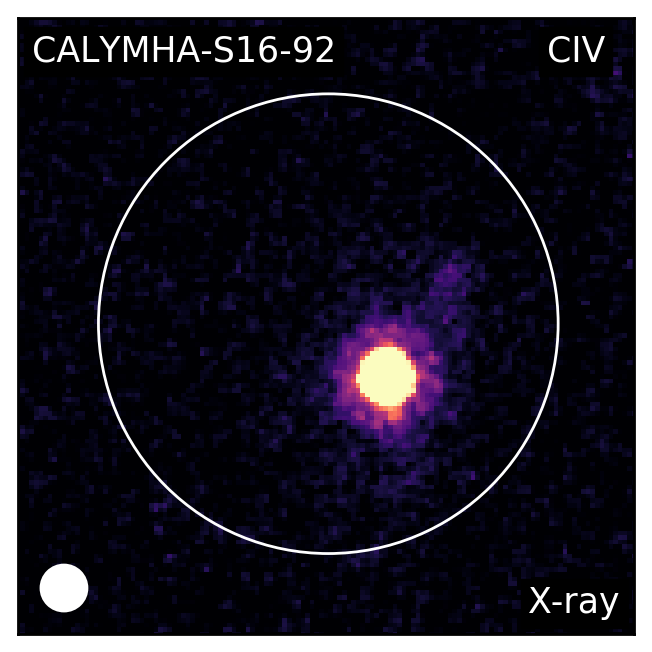}
\includegraphics[trim=0cm 0cm 0cm 0cm, width=0.192\textwidth]{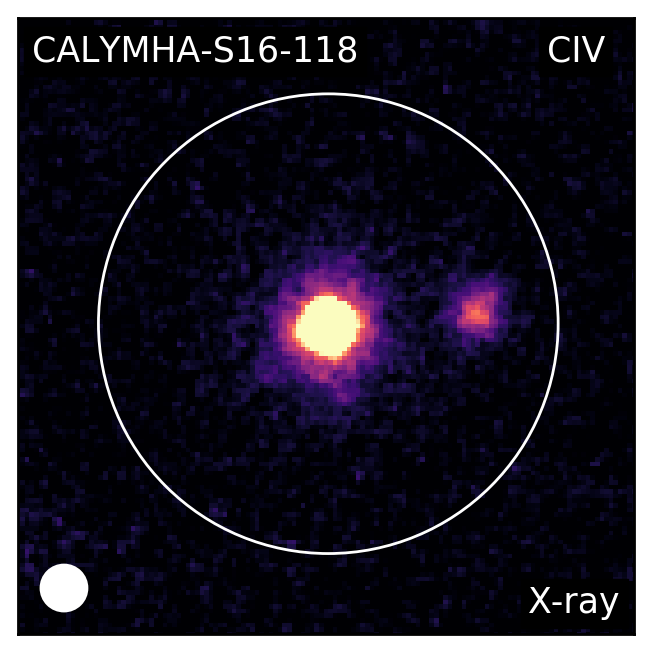}
\includegraphics[trim=0cm 0cm 0cm 0cm, width=0.192\textwidth]{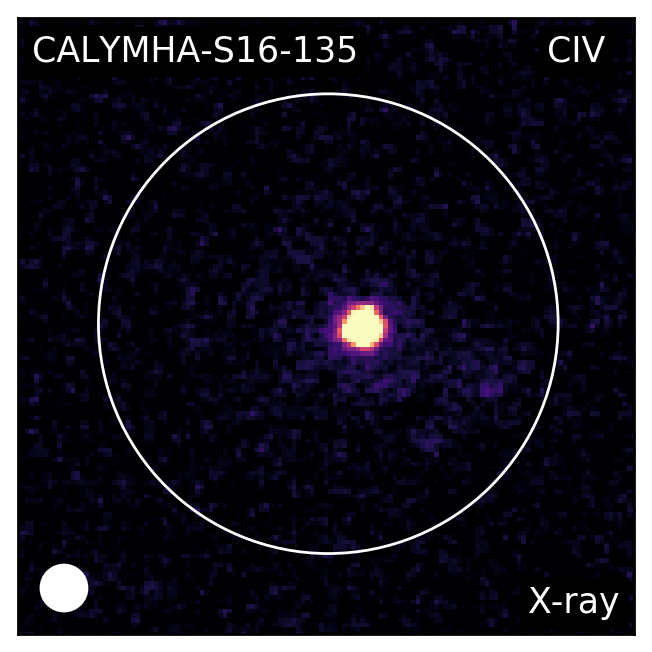}
\includegraphics[trim=0cm 0cm 0cm 0cm, width=0.192\textwidth]{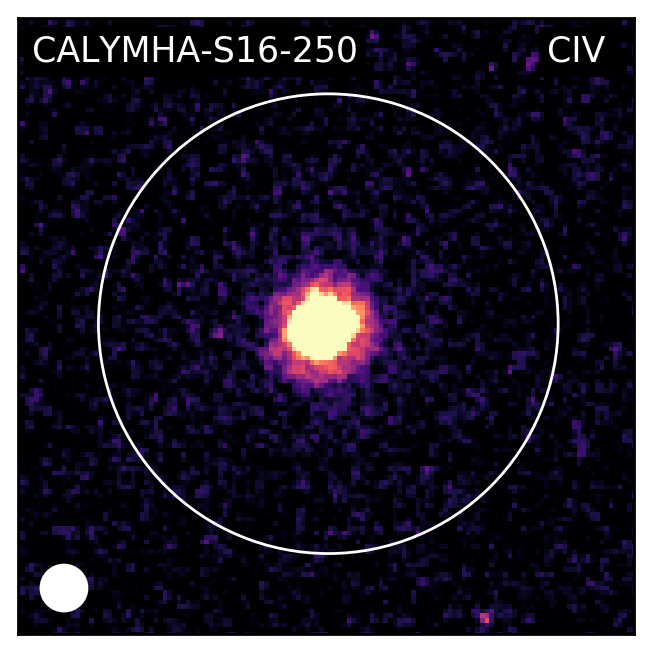}
\includegraphics[trim=0cm 0cm 0cm 0cm, width=0.192\textwidth]{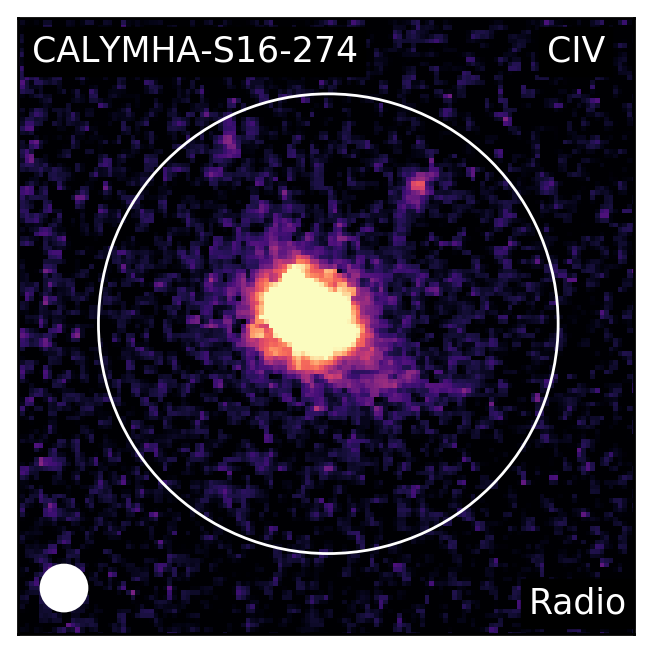}
\includegraphics[trim=0cm 0cm 0cm 0cm, width=0.192\textwidth]{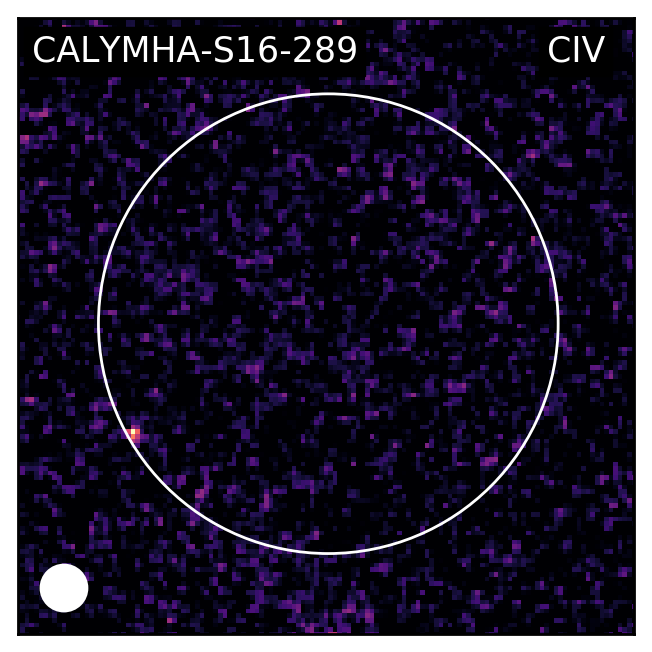}
\caption{Same as Fig.~\ref{fig:HSTCII}, but for \CIV emitters. The cutouts trace $33.9$\,kpc on each side, at the redshift of \CIV ($z\sim1.53$). The large circle represents the $3''$ CALYMHA aperture is marked with the large circle and the small circle is $3$ times the {\it HST} PSF. All the sources are consistent with being point-sources, with their flux contained within the PSF of {\it HST}. Six out of ten sources with {\it HST} coverage have either radio or X-ray counterpart, further supporting a scenario where \CIV sources are power almost exclusively by AGN. Note that all sources with X-ray coverage do have a counterpart. Unlike, \CC or \C emitters, the \CIV line detections are well centred with respect to the {\it HST} and thus more consistent with astrometric errors rather than physical offsets. An AGN powering source for \CIV can readily explain the large EW.}
\label{fig:HSTCIV}
\end{figure*}

Another possibility to explain the large EW are offsets between the main line emission region and the galaxy stellar light (see Figs.~
\ref{fig:HSTCII}, \ref{fig:HSTCIII}, \ref{fig:HSTCIV} and \ref{fig:EWoffset}). Such offsets could be caused by systematic astrometric errors, but can also be caused by a real physical separation in the peaks of the underlying continuum and the line emission. While investigating this avenue with the INT NB data alone is not possible because of the large point spread function (psf, $\sim2$ arcsec), we discuss this in more detail using high resolution {\it HST} data in Section~\ref{sec:morphologies}. As will be shown in Section~\ref{sec:morphologies}, astrometric errors are likely not the cause of the large EW.  

The distribution of rest-frame EW for the \CC emitters extends up to $200$\,{\AA} (as shown in Fig.~\ref{fig:EW}). The chances of all of these high EW sources being interlopers or variable sources is small as explained above. We have also been extremely conservative in our selection and one source with a large $EW_{\rm rest}$ of 84\,{\AA} as measured from our NB data has a  spectroscopic redshift confirming it to be a \CC emitter at $z\sim0.68$ (although this specific source could still be variable).

The rest-frame EW distribution of \C emitters extends up to large values, with an average of $\sim100$\,{\AA} (Fig.~\ref{fig:EW}). These values place our \C sample in a different regime than other samples from the literature. Very recently, \citet{2017ApJ...838...63D} published a spectroscopically-selected \C sample at $z\sim1$ with a median $EW_{\rm rest}$ of $1.3$\,{\AA}. They also compared with results from the literature at redshifts up to $\sim6$ and found that, on average, the \C emitters discovered till now have EW of at most $\sim25$\,{\AA} and are hosted by young, low mass, SF galaxies. The average EW of our \C emitters is also higher than what was found by \citet{2003ApJ...588...65S} in stacks of LBGs, where their strongest Ly$\alpha$ emitters have mean rest-frame \C EW of about $10$ per cent of Ly$\alpha$, or $\sim5.4$\,\AA. The spectroscopically confirmed sources in our \C sample have lower EWs than our average ($14\pm1$, $21\pm9$, $26\pm3$ and $39\pm5$ {\AA} versus the average of $93\pm59$\,{\AA}). Note that both our $z_{\rm spec}$ confirmed \C sources and the literature samples were chosen for spectroscopic followed-up up because of their relatively bright nature. Since sources with spectra are continuum selected this will bias samples towards lower EWs, while our samples are line emission selected, thus finding higher EWs. Hence it is not surprising that the high EW \C sources have not been confirmed with spectroscopy, in the absence of a dedicated, targeted spectroscopic follow-up programme. In addition, as mentioned earlier, any offset between the line emitting region and the stellar light could also explain the large EW we are measuring. Therefore, for continuum selected sources, placing the spectroscopic slit or fibre on the main continuum emitting region does not guarantee the line emission will be captured. Spectroscopy focusing on the line-emitting regions is however crucial to unveil the nature of these emitters and understand the source of the high EW emission.

The \CIV emitters have average rest-frame EWs of $52$\,{\AA}. It is important to note that a significant fraction of these are also spectroscopically confirmed, including a source with a EW of $\sim100$\,{\AA} measured using the NB and BB data (see Fig.~\ref{fig:EW}). The values are consistent within the ranges measured from a large sample of $\sim150$ Type II quasars at $z\sim2-4.3$ by \citet{2013MNRAS.435.3306A}. They measure rest-frame EW ranging from 10 to 100 \,{\AA} for their most secure Type II sample, with an average of about $40$\,{\AA}. Note that the average \CIV line fluxes in \citet{2013MNRAS.435.3306A} are about $(1-1.2)\times10^{-16}$ erg\,s$^{-1}$, which is smaller than our average of $7.7\times10^{-16}$ erg\,s$^{-1}$, but consistent within the spread of the values. Without spectroscopic information, many of our \CIV sources would have been likely Ly$\alpha$ candidates at higher redshifts. While traditionally it has been assumed that large EW emitters are Ly$\alpha$, without clear, secure redshift information the emitter can be misidentified. It thus becomes apparent that \CIV emitters can represent an important contaminating population, as was concluded in \citet{Sobral2017} and in \citet{PaperII}.

\subsection{\textit{HST} morphologies}\label{sec:morphologies}

Ancillary high-resolution space based optical data is crucial in unveiling the nature of our emitters, especially in understanding their colour properties from Sections~\ref{sec:colcol}, \ref{sec:UBcolor} and \ref{sec:beta} and their high EWs (Section~\ref{sec:EW}). For the sources with coverage in {\it HST}, we show their morphologies in Figs.~\ref{fig:HSTCII}, \ref{fig:HSTCIII} and \ref{fig:HSTCIV}. The optical properties of the emitters are listed in Tables~\ref{tab:indivCII}, \ref{tab:indivCIII} and \ref{tab:indivCIV} .

Twelve out of the 16 \CC sources have {\it HST} coverage, revealing a mix of morphology types. Seven \CC emitters have a disky or spiral morphology, while four have a very bright nucleus and a spiral structure, indicative of a Seyfert nature. These Seyfert-like sources are also the most luminous in the emission line and have the lower EW (see Section~\ref{sec:EW}). The possible mixed nature of \CC emitters, some being powered by SF and some by AGN is therefore not only supported by the optical and UV colours of the  emitters (Sections~\ref{sec:colcol}, \ref{sec:UBcolor}, \ref{sec:beta}), but also by the morphologies.

Out of our 24 \C sources, there are 19 with {\it HST} coverage, presenting a wide variety of morphologies. Some \C emitters have a spiral structure, others a disturbed morphology and multiple nuclei, others consisting of 2 or more interacting components. There are also two sources with a UV bright core, indicative of a Seyfert nature, which are also among the brightest in the \C emission line. 

Ten out of 17 \CIV sources have space telescope coverage and they are all point sources even at {\it HST} resolution. The \textit{HST} data support a scenario where \CIV emitters are predominantly quasars: all four sources with {\it Chandra} coverage have a direct X-ray detection, with an additional three sources with a radio detection. We note that the line luminosities of these sources span the $L_{\rm CIV}\sim10^{42.6}-10^{42.9}$ erg s$^{-1}$ range.

\begin{figure}
\centering
\includegraphics[trim=0cm 0cm 0cm 0cm, width=0.479\textwidth]{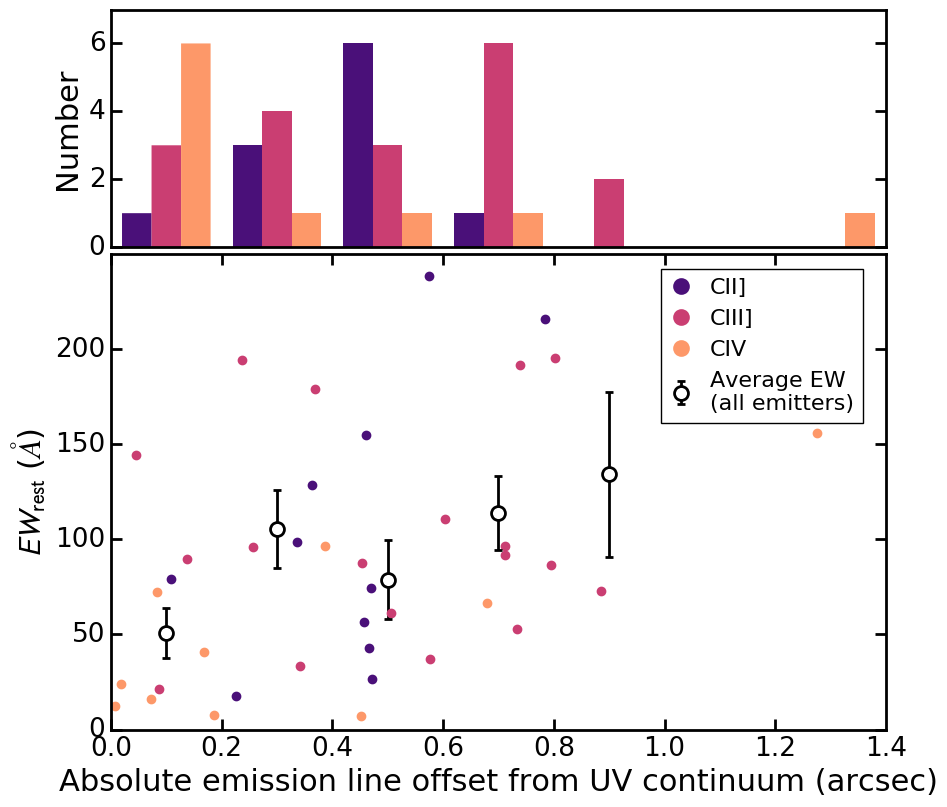}
\caption{Rest-frame EW distribution with respect to offset between the emission line region and the underlying UV continuum. On average, the EW increases with spatial offset, indicating that part of the reason for the high EW is that strong emission line regions are offset from the stellar disk. The EW distribution for \CC and \C emitters peaks at larger spatial offsets, while for \CIV which we find are mainly AGN, the offsets are not as large.}
\label{fig:EWoffset}
\end{figure}

We find that many line emitters, especially \CC and \C are offset from the {\it HST} detection. This is illustrated in Fig.~\ref{fig:EWoffset} where we plot the absolute offset between the emission line regions and the UV continuum against the rest-frame EW. As the distribution of offsets shows, larger offsets are more prevalent for \C emitters, while few large offsets are detected in the \CIV emitter sample. There is also a trend of increasing EW with spatial offset when looking at the population as a whole. It is important to note that sources where the galaxy disk is most offset from the emission line detection in CALYMHA are also among the sources with the largest EW (see discussion of the large EW in Section~\ref{sec:EW}). While there are a few \CC and \C emitters with large EW at small offsets, the bulk of these emitters present offsets larger than 0.2 arcsec. At least for part of the \CC and \C sample, a reason for large $EW_{\rm rest}$ could be the offset between the main line emission region and the location of the bulk stellar light. Therefore, we have tentative evidence that some of the emission line regions can be offset by $1''$ from the brightest UV continuum component, which would result in large $EW$ (see Figs.~\ref{fig:HSTCII} and \ref{fig:HSTCIV}). While the offset can be caused by a physical displacement, a caveat to note is the large PSF of the NB and BB INT observations used to calculate the EW, which can be prone to larger astrometric uncertainties and thus contribute towards the offsets we are measuring. We investigate the offset between CALYMHA sources and the {\it HST} counterparts and found that astrometric errors are in the $0.2-0.3$ arcsec range, which is the expected value for INT observations \citep{Sobral2017}. Therefore, the large offsets found here are most likely of a physical origin, either by the presence of an AGN, and in the case of SF sources, by an offset between the stellar light and the line emitting region. Another option, at least for some of the sources which are not spectroscopically confirmed, would be that they are in fact higher redshift, lensed Ly$\alpha$ emitters. This may explain the offsets as well as the large EW.

\subsection{\textit{Chandra} X-ray and VLA radio counterparts}\label{sec:ancillary}

While no emitters have direct FIR detections (and thus none can be extremely star-forming at the $1000$ M$_\odot$\,yr$^{-1}$ level), some sources do have an X-ray or radio counterpart, which we discuss here and summarise in Table~\ref{tab:summary}. Five \CC have {\it Chandra} coverage and 13 are covered by radio data. Nine \C sources are covered with {\it Chandra} and 22 with radio coverage, while 4 \CIV are covered with X-ray and 13 with radio.

In the case of \CC, the results from the optical morphologies and colours are supported by the X-ray data: the disk galaxies with {\it Chandra} coverage do not have an X-ray counterpart, while the Seyferts have counterparts. None of the \CC sources is active in the radio (out of 13 with coverage). The {\it HST} data reveals that some of the \CC emitters have a spiral-like morphology. In that canse any radio emission will be powered by SF, which is low compared to AGN-powered radio emission. Additionally, only 5 per cent of Seyferts are radio-loud. Therefore, the lack of radio detection for \CC emitters is not surprising given the depth of the radio observations. 

As mentioned in Section~\ref{sec:morphologies}, the \C sources have a range of morphologies. One of the two sources with a UV bright core, indicative of a Seyfert nature, has coverage in {\it Chandra} and VLA and possesses an X-ray and radio detection. As expected for normal SF galaxies, none of the other \C sources have radio or X-ray counterparts. A detection would imply very large SFRs of $\sim1000$\,M$_{\odot}$\,yr$^{-1}$.

The radio and \textit{Chandra} data for \CIV, in agreement with \textit{HST}, support a scenario where \CIV emitters are predominantly quasars: all four sources with {\it Chandra} coverage have a direct X-ray detection, with an additional three sources with a radio detection. 

In conclusion, the X-ray and radio data suggest that \C is mostly powered by SF. \CIV and some \CC emitters are active in the X-ray, indicating they are young, actively accreting in the radiatively driven, quasar mode. Only a few \CIV and no \CC have radio detections, which would be indicative of a more evolved AGN in the mechanical, radio-loud phase. 

\subsection{Average SFRs and BHARs}

\subsubsection{{\it Chandra} stacking}

To evaluate the level of activity of the central supermassive black hole in our emitters, we can estimate the black hole accretion rate (BHAR) which can be derived from X-ray data. We follow the method described in \citet{2017MNRAS.464..303C} to measure the average BHAR for the \CC, \C and \CIV emitters by stacking sources with {\it Chandra} coverage. 

We make a $20''\times20''$ cutout around the position of each source and take the mean to obtain an average image which is used to calculate the average flux of the emitters. We establish a stacking radius of 2 arcsec and convert the counts inside to flux by following the indicated procedure in \citet{2009ApJS..184..158E}, with a photon index $\Gamma$ of 1.4 and Galactic absorption of $2.7\times10^{20}$ cm$^{-2}$. We convert the flux $F_{\rm X}$ to an X-ray luminosity $L_{\rm X}$ using:
\begin{equation}
L_{\rm X} = 4 \pi D_{\rm L}^2 F_{\rm X} (1+z)^{\Gamma-2}\quad ({\rm erg\,s}^{-1}), 
\end{equation}
We obtain the bolometric luminosity by multiplying by 22.4 \citep{2013ApJ...765...87L}. The BHAR is obtained by:
\begin{equation}
BHAR = \frac{(1-\epsilon)L_{\rm bol}}{\epsilon c^2}\quad ({\rm M}_\odot\,{\rm yr}^{-1}),
\end{equation}
where $\epsilon=0.1$ is the accretion efficiency \citep[following][]{2013ApJ...765...87L} and $c$ is the speed of light. Errors on both $L_\mathrm{X}$ and the BHAR are calculated by taking the standard deviation of the distribution obtained from a bootstrapping analysis of the data.

\begin{figure}
\centering
\includegraphics[trim=0cm 0cm 0cm 0cm, width=0.479\textwidth]{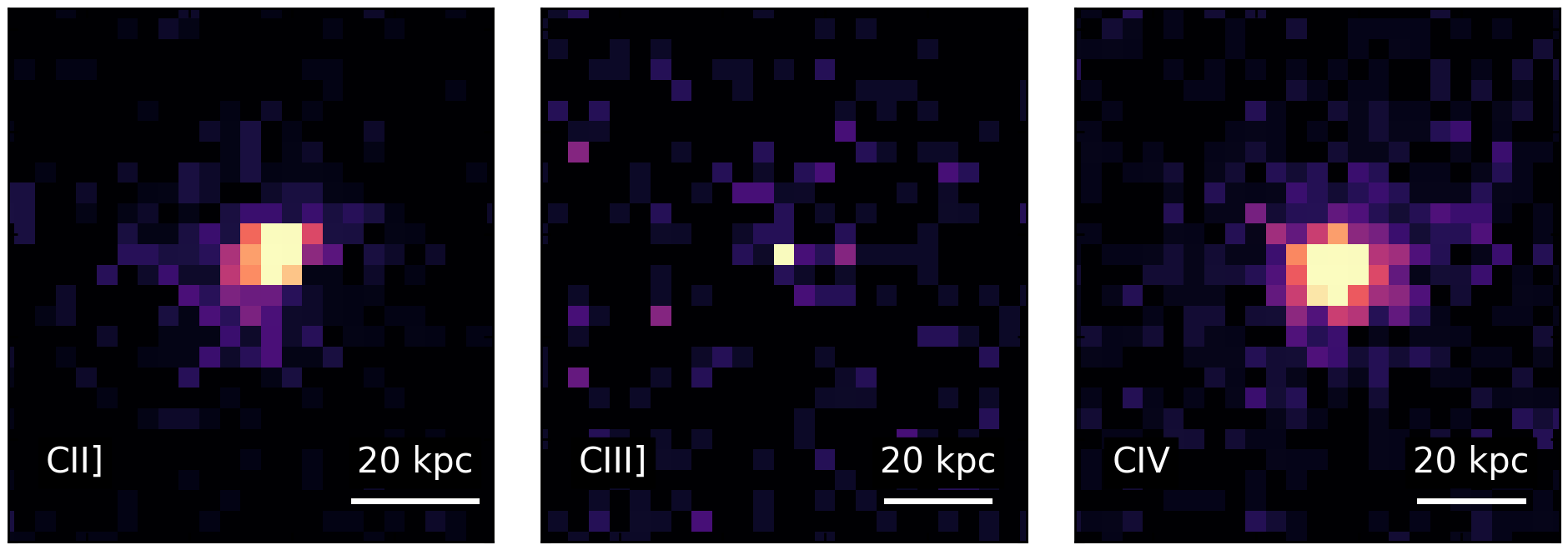}
\caption{The X-ray stack for sources with coverage in COSMOS with {\it Chandra}. The \CC and \CIV detections are consistent with AGN activity with large X-ray luminosities of $10^{43.5}$ and $10^{44.4}$ erg\,s$^{-1}$, respectively. The much fainter \C, with an average X-ray luminosity of $10^{42.8}$ erg\,s$^{-1}$, is likely dominated by SF galaxies with a very minor AGN contamination, consistent with our results. The images are on the same colour scale.}
\label{fig:stack}
\end{figure}

\begin{table*}
\caption{Summary of the optical, X-ray and radio properties of \CC, \C and \CIV emitters. The full details on individual sources can be found in Tables~\ref{tab:indivCII}, \ref{tab:indivCIII} and \ref{tab:indivCIV}.}
\centering
\begin{tabular}{l c c P{5cm} c c c c c c}
\hline\hline
Line & & \multicolumn{2}{c}{\hspace{-20pt}\it HST} & &  \multicolumn{2}{c}{\it Chandra} & &  \multicolumn{2}{c}{Radio $1.4$ GHz} \\ \cline{3-4} \cline{6-7} \cline{9-10} 
& & coverage & morphology & & coverage & counterpart & & coverage & counterpart \\
\hline
\CC  & & 12/16 & 4 bright-core+disk, 7 disky, 1 interacting & & 5/16 & 2/5 & & 13/16 & 0/13  \\
\C   & & 19/34 & 2 bright-core+disk, 8 disturbed/interacting, 7 diffuse/spiral, 2 compact  & &  9/34 & 1/9 & & 22/34 & 1/22 \\
\CIV & & 10/17 & 10 point sources & & 4/17 & 4/4 & & 12/17 & 3/14 \\ \hline
\end{tabular}
\label{tab:summary}
\end{table*}

\subsubsection{FIR and radio stacking}

We use FIR and radio data to obtain an average SFR in our sources.

To obtain an estimation of the dust-obscured SFR from the FIR emission, we start by mean stacking our sources using the \textit{Herschel} bands at 100, 160, 250, 350 and 500 $\mu$m (PACS and SPIRE instruments). Aperture corrections were applied for the PACS 100 and 160\,$\mu$m bands as specified in the PACS PEP release notes. In the SPIRE 250, 350, 500\,$\mu$m, the fluxes were taken from the peak value in each stack. Note that we do not have any detections and can only provide upper limits. We estimate the FIR luminosities by fitting modified black body templates \citep[using the SWIRE template Library,][]{2007ApJ...663...81P} to the upper limits and integrating the best fit between 8 and 1000\,$\mu$m. We use the total FIR luminosity $L_{\rm FIR}$ to compute the SFRs, assuming a \citet{2003PASP..115..763C} IMF by using:
\begin{equation}
SFR_{\rm FIR} = 2.5 \times 10^{-44} L_{\rm FIR} \quad  {\rm M}_\odot\,{\rm yr}^{-1}.
\end{equation}
We find no detections in any of the stacks and thus can place upper limits on the SFRs. The results of the stacking do not change if we use all sources with coverage or only the sources with {\it Chandra} coverage.

Radio at $1.4$ GHz can be used as a dust-free SF indicator on timescales longer than FIR or emission lines. The radio stacking procedure is the same as the X-rays. To convert the radio luminosities $L_{\rm radio}$ to SFRs, we use the conversion determined by \citet{2001ApJ...554..803Y}, adapted to a \citet{2003PASP..115..763C} IMF:
\begin{equation}
SFR_{\rm radio} = 3.18 \times 10^{-22} L_{\rm radio} \quad {\rm M}_\odot\,{\rm yr}^{-1}.
\end{equation}

Note that in the case of radio, unlike FIR, a direct detection is caused by an AGN rather than SF. We find that if we remove all detections, our radio stacks all provide non-detections in the same range as the FIR ones.

\begin{figure*}
\centering
\includegraphics[trim=0cm 0cm 0cm 0cm, width=0.699\textwidth]{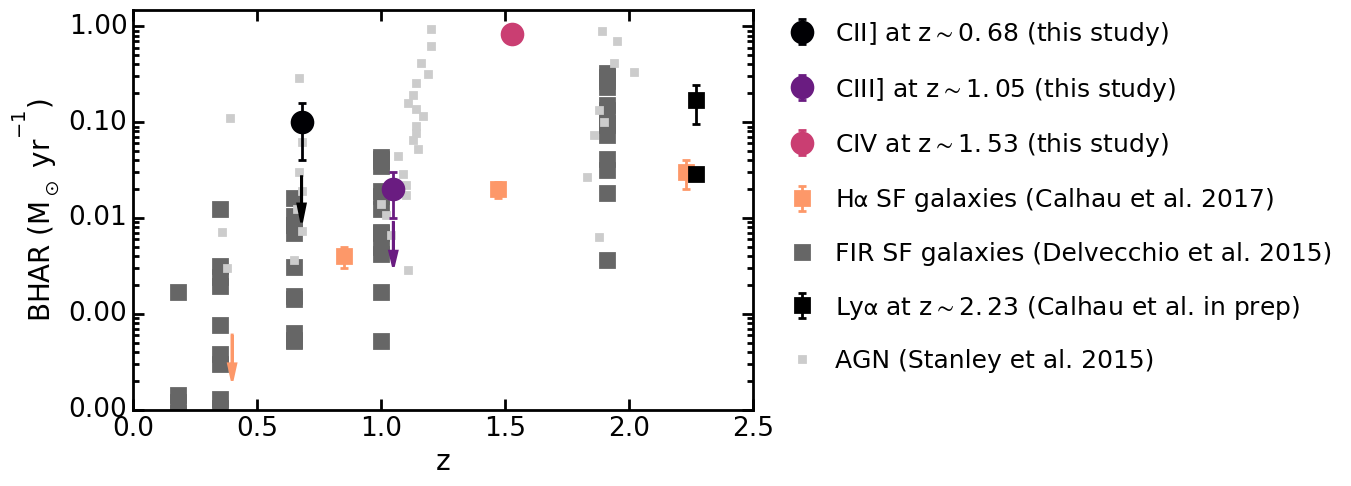}
\caption{The black hole accretion rate ($BHAR$) for the stacked \CC, \C and \CIV sources, obtained by converting their average X-ray luminosities into accretion rates. All \CIV sources are directly detected in the X-ray. Upper limits for \CC and \C stacks are given with direct detections removed. \CC emitters have a high average $BHAR$, similar to AGN. \C emitters have $BHAR$ similar to H$\alpha$ selected, main sequence SF galaxies. \CIV emitters have extremely high accretion rates, consistent with bright quasars. We also plot the $BHAR$ for H$\alpha$ selected, SF galaxies \citep{2017MNRAS.464..303C}. Overplotted in dark grey are main-sequence, FIR selected SF galaxies from \citet{2015MNRAS.449..373D}. The different values are stacks for galaxies of different masses and SFRs. The light grey points represent AGN from \citet{2015MNRAS.453..591S} stacked in different luminosity bins. We also plot the result for Ly$\alpha$ emitters at $z\sim2.23$ from Calhau et al. (in prep), with and without the direct X-ray detections included in the stacks.}
\label{fig:BHAR}
\end{figure*}

\subsubsection{Stacking results}

The results of our X-ray, FIR and radio stacking analysis are listed in Table~\ref{tab:Xray}. The stacked X-ray image can be found in Fig.~\ref{fig:stack}, which shows clear detections for \CC and \CIV and a very faint/non detection for \C. We plot our $BHAR$ results and compare with similar measurements obtained for H$\alpha$ and FIR selected SF galaxies, Ly$\alpha$ emitters and AGN \citep[][Calhau et al. in prep]{2015MNRAS.449..373D, 2015MNRAS.453..591S, 2017MNRAS.464..303C} in Fig.~\ref{fig:BHAR}. Note however the low number statistics, driven by the small area of the publicly available {\it Chandra} images.

The stacking analysis indicates that \CC emitters are bright in the X-rays and are consistent with AGN activity ($BHAR=0.10\pm0.06$ $M_\odot$\,yr$^{-1}$), as revealed by Fig.~\ref{fig:BHAR}. \CC emitters accrete matter onto their BH at a much higher rate than SF galaxies. The \CC emission, as suggested, for example, by \citet{2002RMxAC..13..155B} could then be in part powered by shocks near the BH.

Once the only source with an X-ray detection is removed, \C emitters have an upper limit X-ray luminosity consistent with SF activity and a low BHAR ($0.02\pm0.01$ $M_\odot$\,yr$^{-1}$), similar to what is measured from H$\alpha$ and FIR selected SF galaxies \citep{2015MNRAS.449..373D,2017MNRAS.464..303C}. 

The average X-ray luminosity of the \CIV sources is very high, $\sim10^{44.4}$ erg\,s$^{-1}$, supporting an active nucleus which accretes matter at a rate of $\sim0.83$ $M_\odot$\,yr$^{-1}$. This value is well above the rate expected for main sequence SF galaxies, lying in the AGN region of Fig.~\ref{fig:BHAR}. This is consistent with the optical properties of the \CIV emitters discussed in previous sections.

There are no detections in the FIR bands, so we are only able to provide $3\sigma$ SFR upper limits of $15$, $75$, $141$ M$_\odot$\,yr$^{-1}$ for \CC, \C, \CIV, respectively. Note that the different values are mostly driven by the different redshifts of the three types of emitters. The FIR non-detections are mostly caused by the relatively low number of sources in each stack compared to the detections from \citet{2017MNRAS.464..303C}, who stacked about ten times more sources than included in this analysis. In order to further test this, we stack the H$\alpha$ emitters at $z\sim0.84$ in the FIR, using the same number of sources as our \C emitters and find no detection. We therefore conclude that \C emitters are fully consistent with being similar to H$\alpha$ selected samples at a similar redshift, including the fact that, as seen before, they are slightly redder than LBGs.

Similarly, we do not have a detection in the radio stacks after removing direct detections (which are caused by AGN, not SF). These upper limits, as for the FIR, are not very stringent ($\sim17,35,115$ $M_\odot$\,yr$^{-1}$ for \CC, \C and \CIV respectively). The range of typical SFR from H$\alpha$ emission at $z\sim0.8-1.5$ is in the range $8-16$ $M_\odot$\,yr$^{-1}$ \citep{2013MNRAS.428.1128S}, well below our FIR and radio depths.

\subsubsection{UV SFRs}

We also attempt to estimate the SFRs of \CC, \C and \CIV emission from the restframe UV emission. While UV emission is produced by young, massive stars, thus indicating recent SF episodes, this SFR estimator can be heavily affected by the dust content of the host galaxy. As well, in the cases of galaxies hosting an AGN, accretion can also produce significant UV, making it harder to disentangle the contribution from SF.

We estimate the dust extinction $A_{\rm UV}$ in the restframe UV by using the dust extinction law from \citet{2000ApJ...533..682C}:
\begin{equation}
\label{eq:dust}
A_{\rm UV} =  2.31 \beta + 4.85,
\end{equation}
where $\beta$ is the UV restframe slope. In principle this relation has been defined for $\beta$ measured between $1300$ and $2600$\,{\AA}. However, since these wavelengths are not available directly, we use the $\beta$ as measured from our data (see Section~\ref{sec:beta}) with the warning that we are tracing slightly different rest-frame wavelengths which can introduce uncertainties in the dust attenuation estimates. 

We obtain the $SFR_{\rm UV}$ from the UV data using the conversion presented in \citet{1998ARA&A..36..189K}, corrected for the  \citet{2003PASP..115..763C} IMF:
\begin{equation}
\label{eq:SFRUV}
SFR_{\rm UV} = 8.8 \times 10^{-29} L_{\rm UV} \quad {\rm M}_\odot\,{\rm yr}^{-1},
\end{equation}
where:
\begin{equation}
L_{\rm UV}  = 4 \pi 9.523 \times 10^{38} \times 10^{-0.4 (M_{\rm UV, corr}+48.6)} \quad {\rm erg\,s^{-1}\,Hz},
\end{equation}
where $M_{\rm UV, corr}$ is the absolute UV magnitudes corrected for dust attenuation $A_{\rm UV}$.

We list the average and median SFRs for \CC and \C in Table~\ref{tab:Xray}. We would like to point out that these values depend on a number of assumptions and can be affected by uncertainties. The $\beta$ slope is not measured exactly at the wavelengths defined in \citet{2000ApJ...533..682C}. A large unknown is the proportion of UV emission that can actually be attributed to SF, and how much to other support such as AGN activity. To alleviate this as much as possible, we remove sources which we were able to classify as AGN from our ancillary data. In the case of \CIV emitters, which we classified predominantly as quasars, we do not report an UV based SFR. As discussed in the paper, the properties of \CIV emitters are consistent with a population of AGN/quasars, especially for bright sources ($M_{\rm UV}\sim-22$).  Therefore most of the UV for \CIV sources comes from accretion precesses onto the central BH.

We note that for both \CC and \C, the mean SFR is much larger than the median value, indicating a few outliers with very large SFR bias the measurements. 

The $SFR_{\rm UV}$ values for \CC emitters are typically lower than those of \C emitters, with averages a factor of 25 lower and medians a factor of 10 lower. This supports the scenario discussed in previous sections of the paper, where \CC emitters are most likely hosted in SF galaxies and/or AGN.  The average SFR for \C emitters is consistent with values obtained in stacks of typical, H$\alpha$-selected SF galaxies, using FIR as SF tracer. \citet{2017MNRAS.464..303C} find a values of $13$ M$_\odot$\,yr$^{-1}$ at $z\sim0.85$ and $32$ M$_\odot$\,yr$^{-1}$ at $z\sim1.5$. Nevertheless, this comparison should be made with caution given the difficulty in comparing different SFR estimators.

\renewcommand{\arraystretch}{1.2}
\begin{table*}
\centering
\caption{Average X-ray luminosity and BHAR, as resulting from a stacking analysis of the \CC, \C and \CIV emitters (with $z_{\rm spec}$ or $z_{\rm phot}$), including direct detections. While \CC and \CIV emitters are consistent with AGN activity, \C values place in the SF galaxy regime. Note that removing the direct X-ray detections results in non-detections in the stacks because of the very low numbers of sources left. We also present SFR upper limits obtained from FIR and radio data. Note that individual detections are removed from the radio stack as a direct radio detection will be caused by an AGN rather than SF (which is the purpose of our radio stacks). We also report SFR measurements from the UV, assuming all UV radiation is coming from SF. We removed sources where ancillary data strongly indicated the presence of an AGN. Therefore, we report no average UV SFR for \CIV emitters.}
\begin{tabular}{l c c c c c c c c c c c}
\hline\hline
Line  & $\log L_\mathrm{X}$   & $BHAR$  & No sources & $SFR_{\rm FIR}$ $3\sigma$ upper limit & $SFR_{\rm radio}$ $3\sigma$ upper limit & Mean $SFR_{\rm UV}$ & Median $SFR_{\rm UV}$ \\ 
     &  (erg\,$s^{-1}$) & ($M_\odot$\,yr$^{-1}$)&  &  ($M_\odot$\,yr$^{-1}$) &  ($M_\odot$\,yr$^{-1}$) & ($M_\odot$\,yr$^{-1}$) & ($M_\odot$\,yr$^{-1}$) \\ \hline             
\CC & $43.46\pm0.56$ & $0.10\pm0.06$ & 5 & \phantom{0}$<15$ & \phantom{0}$<17$ & $7\pm10$ & $2\pm13$  \\ 
\C  & $42.80\pm0.37$ &  $0.02\pm0.01$ & 9 &  \phantom{0}$<75$ & \phantom{0}$<35$ & $120\pm345$ & $24\pm430$ \\ 
\CIV &  $44.37\pm0.08$ & $0.83\pm0.12$ & 4  & $<141$ & $<115$ & -- & -- \\ \hline 
\hline
\end{tabular}
\label{tab:Xray}
\end{table*}
\renewcommand{\arraystretch}{1.1}

\section{Discussion}\label{sec:discussion}

We investigate the nature of \CC, \C and \CIV line emitters by jointly interpreting their emission line properties with broad band optical colours as well as ancillary space-based optical and X-ray data. 

\subsection{\CC emitters $z\sim0.68$}

\CC emitters at $z\sim0.68$ have optical colours spanning across the colour space of SF galaxies at their redshift into a regime more typical of SF sources at higher cosmic distances. Such unusual optical colours may indicate that while part of the \CC emitter population is powered by SF, at least some sources host an AGN, in agreement with luminosity function and cosmic ratio results from \citet{PaperII}. This is also supported by the {\it HST} morphology of \CC sources: while some sources have spiral-like morphologies, part of the population, particularly the sources with atypical colours, have a bright nucleus accompanied by a disk. While the bright nucleus could be indicative of a nuclear starburst surrounded by a stellar disk, this scenario is not supported by the X-ray data which reveals that these \CC emitters have BHAR rates consistent with an AGN at their core. \CC emitters have relatively steep $\beta$ slopes, which indicate a relatively blue UV continuum. This is also supported by the lack of a detection in the FIR stacks of these sources. \CC emitters also have high rest-frame EWs, which could be caused by either the offset we observe between the main stellar disk and the line emitting region, or by AGN activity.

Traditionally, it was thought that \CC emission is mostly triggered in shocks around AGN \citep{2000A&A...362..519D, 2000MNRAS.311...23B, 2002RMxAC..13..155B}, however this resulted from studies of this line being done almost exclusively in bright, active galaxies. Overall, our results indicate that \CC emitters may actually powered in part by blue, SF galaxies at lower luminosities and in part by Seyfert-like sources with young, actively-accreting AGN at brighter line luminosities. 

\subsection{\C emitters at $z\sim1.05$}

Most \C emitters at $z\sim1.05$ have optical colours consistent with SF galaxies at their redshift. Furthermore, their low average BHAR is also consistent with SF galaxies. The $\beta$ UV slopes of \C emitters are consistent with H$\alpha$ selected galaxies, indicating \C emission might be suitable avenue for selecting a wide range of SF galaxies. \C sources are also mostly X-ray and radio quiet, as expected for SF galaxies. Most emitters have disky, clumpy or disturbed optical morphologies, indicative of either single or interacting SF galaxies. The offset optical disk or the clumpy/disturbed morphologies might explain the large EW we find for some \C emitters. The few direct X-ray and radio detections ($<10$ per cent) correlate with a bright nucleus plus a disk optical morphology, which indicates that the brightest \C emitters host an AGN at their core. 

Traditionally \C emission was thought to originate from AGN. More recent results focusing on SF galaxies find that $\sim20$ per cent of local galaxies have the \C feature and the strongest \C emission is fostered at low metallicities \citep{2015ApJ...814L...6R}. Part of their sample are Wolf-Rayet galaxies, indicating recent (few Myrs), short-lived ($<1$ Myr) strong starbursts. At higher redshifts ($1.5-3$), \citet{2014MNRAS.445.3200S} find that almost all of their \C emitters also possess strong Ly$\alpha$. \citet{2016MNRAS.456..485S}, in agreement with \citet{2016ApJ...833..136J}, showed that the presence of binary stars can increase the ionising flux by up to $60$ per cent in low metallicity ($<0.3Z_{\odot}$) environments, which could also lead to a boost in \C production. 

While these works indicate that strong \C is produced in relatively low-metallicity SF galaxies, our results strongly support a scenario where \C emitters trace a more general population of SF galaxies at $z\sim1$, with very few bright emitters being powered by AGN. This is similar to what is found in studies of Ly$\alpha$ and H$\alpha$ emitters at $z>2$ \citep{2015MNRAS.451..400M, 2016ApJ...823...20K, Sobral2017, Matthee2017}. However, as shown in \citet{PaperII}, at fixed UV magnitude of line flux, the number density of \C emitters is substantially lower than H$\alpha$ or LBG SF galaxy populations, indicating only a fraction of galaxies have strong \C. The average line ratios imply that \C is $\sim20$ times fainter than H$\alpha$ and $\sim4$ times fainter than observed Ly$\alpha$
\citep{PaperII}.

An open question relates to the large $EW_{\rm rest}$ we measure for bulk of the \C population. According to photoionisation modelling, the largest EWs that can be reached with star formation are of the order of $25$\,{\AA} \citep{2016ApJ...833..136J}. As discussed earlier in the paper, the large EW cannot be solely attributed to systematic errors in the EW estimation or effects such as variability. Whether the discrepancies between our observations and the modelling can be attributed to spatial offsets between emission line regions we cannot ascertain for sure with the current data. Spectroscopic follow-up will reveal the detailed physics of the \C emitters, as well as confirm or dismiss the large EW measured here.
 
\subsection{\CIV emitters $z\sim1.53$}

The direct detections in the X-ray, the high accretion rates characteristic for bright AGN, the universal optical quasar morphologies, the unusual optical colours and blue UV slopes and large rest-frame EW, all reveal that \CIV is produced in intense radiation fields characteristic to young, virtually dust-free actively accreting quasars. The number distribution of \CIV emitters as function of line luminosity also indicates they are primarily a quasar population, being well described by a power-law \citep{PaperII}.

In a sample of lensed, low mass galaxies at $z\sim1.5-3$, \citet{2014MNRAS.445.3200S} find ubiquitous joint detections of \CIV and \C, which require young stellar populations, increased electron temperatures and an enhanced ionising output from metal poor gas and stars. While the modelling is still uncertain, their work therefore indicates that the \CIV emission is mainly driven by young stars, without particular need for an AGN contribution. 

Our results however, seem to support a scenario where the bulk of the \CIV is produced in galaxies hosting powerful AGN at least at $z\sim1.5$. One important aspect to note is that the rest-frame UV magnitudes of the sources from \citet{2014MNRAS.445.3200S} are similar to ours, so a comparison is appropriate from this point of view. However, the galaxies in  the \citet{2014MNRAS.445.3200S} sample are at a higher redshift and also span on average lower EW values ($<15$\,{\AA} for the Stark sample versus $34\pm58$\,{\AA} for our sample).

\section{Conclusions}\label{sec:conclusion}
We characterise the properties \CC, \C and \CIV emitters at $z\sim0.68,1.05,1.53$. These sources were selected from the first blind, statistical study of Carbon emitters, uniformly selected in the COSMOS and UDS field as part of the CALYMHA survey. In this paper, we focus on the UV, optical, X-ray, FIR and radio properties of these emitters. Our main results are:
\begin{itemize}
\item \CC emitters at $z\sim0.68$ have disky structure or bright nucleus with a disk morphology (Fig.~\ref{fig:HSTCII}) and blue UV colours (Figs. \ref{fig:UB}, \ref{fig:beta}). Some of the brightest line emitters are strong X-ray sources (Fig.~\ref{fig:stack}), with black hole accretion rates (Fig.~\ref{fig:BHAR}) characteristic of AGN. Our results therefore indicate that \CC traces a mixed population of SF galaxies and AGN, especially at the brightest line luminosities. For the brighter emitters, our results are in accordance with the theoretical expectation that \CC emission is triggered by shocks around AGN.
\item \C emitters at $z\sim1.05$ in our study have typical optical colours for SF galaxies (Fig.~\ref{fig:colcol}) and morphologies indicative of either isolated or interacting galaxies (Fig.~\ref{fig:HSTCIII}). Their relatively blue rest-frame UV and optical colours (Figs. \ref{fig:iz}, \ref{fig:beta}) and low black hole accretion rates (Fig.~\ref{fig:BHAR}) are consistent with general populations of SF galaxies. 
\item The point-like {\it HST} morphologies (Fig.~\ref{fig:HSTCIV}), the strong X-ray detections implying high accretion rates (Figs.~\ref{fig:stack}, ~\ref{fig:BHAR}) and the blue optical and UV colours (Figs.~\ref{fig:UB}, \ref{fig:beta}) indicate that \CIV emitters are almost universally blue/non-dusty quasars. 
\item We discover a large, previously-undetected population of \CC, \C and \CIV emitters with large rest-frame EW, of above $50$\,{\AA}, potentially up to 200\,{\AA} (Fig.~\ref{fig:EW}). In stark contrast, the typical EW of continuum selected \C samples from the local Universe up to $z\sim6$ is $\lesssim25$\,{\AA}. These large EW might be caused by emission line regions offset from the stellar disk or because of the AGN nature of some of the sources.
\end{itemize}

While historically thought to originate in AGN, our study revealed some interesting properties of \CC, \C and \CIV emitters at $z\sim0.7-1.5$. While \CIV emitters are almost universally quasars, \C emission is consistent with being produced in a wide variety of SF galaxies and \CC emission may trace either an AGN or SF. Further spectroscopy is crucial however to unveil their powering source and the physics of the large EW we measure in a fraction of the population. In the companion paper, \citet{PaperII}, we further explore the samples and derive luminosity functions and cosmic average line ratios.

\section*{Acknowledgements}

We would like to thank the anonymous referee for her/his valuable input that helped improve the clarity and interpretation of our results. DS acknowledges financial support from the Netherlands Organisation for Scientific research (NWO), through a Veni fellowship. IO acknowledges support from the European Research Council in the form of the Advanced Investigator Programme, 321302, {\sc cosmicism}. CALYMHA data is based on observations made with the Isaac Newton Telescope (proposals 13AN002, I14AN002, 088-INT7/14A, I14BN006, 118-INT13/14B, I15AN008) operated on the island of La Palma by the Isaac Newton Group in the Spanish Observatorio del Roque de los Muchachos of the Instituto de Astrof{\'i}sica de Canarias. Also based on data products from observations made with ESO Telescopes at the La Silla Paranal Observatory under ESO programme IDs 098.A-0819 and 179.A-2005. We are grateful to E. L. Wright and J. Schombert for their cosmology calculator. We would like to thank the authors of NumPy \citep{numpy}, SciPy \citep{scipy}, Matplotlib \citep{matplotlib} and AstroPy \citep{astropy} for making these packages publicly available. This research has made use of the NASA/IPAC Extragalactic Database (NED) which is operated by the Jet Propulsion Laboratory, California Institute of Technology, under contract with the National Aeronautics and Space Administration. This research has made use of NASA's Astrophysics Data System. This research has made use of the VizieR catalogue access tool, CDS, Strasbourg, France. The original description of the VizieR service was published in \citet{2000A&AS..143...23O}. This research has made use of ``Aladin sky atlas" developed at CDS, Strasbourg Observatory, France \citep{2000A&AS..143...33B,2014ASPC..485..277B}. The CALYMHA catalogue used for this study is publicly available from \citet{Sobral2017}.

\bibliographystyle{mn2e.bst}

\bibliography{CarbonEmittersI}

\appendix

\section{Effect of variability on the measured EW}\label{sec:variability}

Since the NB and BB observations were taken in different years, it is possible that any variable source (such as an AGN) is falsely selected as a line-emitter or has an over- or under- estimated EW. This is because, on such timescales, the continuum UV magnitudes of AGN may change by $\sim1$ magnitude \citep[e.g.][]{1997ARA&A..35..445U, 2003AJ....126.1217D, 2005A&A...443..451F, 2011ApJ...731...97S, 2017MNRAS.465..302M}. Thus, there is the possibility that some of the large EW sources are not line-emitters, but variable sources caught in a high state in the NB, but in a low state in the BB, thus mimicking a colour excess. Another possibility is that some of the EW are overestimated because of variability.

We empirically estimate the number of variable sources that may contaminate our sample as follows. If such variable sources exist, we expect not only to find them in the sample of line-emitters, but also in samples of absorbers (with negative excess). Therefore, we select galaxies with statistically significant negative excess \citep[from][]{Sobral2017}. Among these, we focus on X-ray AGN by matching negative excess sources with the catalogue of X-ray detected sources from \citet{2012ApJS..201...30C}, resulting in 13 matches (all with available spectroscopic redshifts). We remove four sources for which the absorption feature corresponds to Ly$\alpha$ forest between $912-1216$\,{\AA}, as well as two sources for which the absorption feature can be explained by C{\sc iv} absorption. The remaining 7 AGN are at redshifts $0.9<z<2.0$, and no obvious absorption feature is present at the wavelength traced by the NB. Thus, we assume that these 7 AGN are in fact variable sources that mimic an absorption feature. These AGN have randomly distributed excesses between -0.4 and -1.1. If we would assume that such variable sources with such excess exist, we would expect 7 line-emitters to be variable sources with $EW_{\rm obs} \approx 30-110$\,{\AA}. Therefore, while a few of the large EW sources might be variable AGN, variability cannot fully explain the distribution of EW. One possibility we cannot rule out is variability with a smaller magnitude ranges ($-0.2$ to $0.2$ mag), because of photometric errors. Therefore while some EW may be boosted, in some cases, the measured EWs will be lower than in reality. Spectroscopic follow-up is required to completely verify some of the highest EWs.

\section{Properties of individual \CC, \C and \CIV emitters}

In Tables~\ref{tab:indivCII}, \ref{tab:indivCIII} and \ref{tab:indivCIV} we list the individual properties of \CC, \C and \CIV emitters, respectively. We give the coordinates of the sources, their optical {\it HST} morphology, possible X-ray and radio counterparts, as well as line luminosity, rest-frame EW ($EW_{\rm rest}$) and the observed $(U-B)_{\rm obs}$ colour.

\begin{table*}
\begin{threeparttable}
\caption{Properties of the $z\sim0.7$ \CC emitters: coordinates, spectroscopic redshift information availability, {\it HST} optical morphologies and possible X-ray and radio counterparts. We also list the emitters' line luminosity, rest-frame $EW_{\rm rest}$ and observed $(U-B)_{\rm obs}$ colour. The error on the $(U-B)_{\rm obs}$ colour was calculated assuming the maximal magnitude errors of $0.05$. A fits version of this table can be found online.\vspace{-5pt}}
\centering
{\scriptsize
\begin{tabular}{l c c c c c c c r@{$\pm$}l c }
\hline\hline
CALYMHA ID & \multicolumn{1}{c}{RA}   & \multicolumn{1}{c}{DEC} & $z_{\rm spec}$ & {\it HST}  & {\it Chandra}    & Radio $1.4$ GHz & $\log L_{\rm CII]}$ & 
\multicolumn{2}{c}{$EW_{\rm rest}$} & \multicolumn{1}{c}{$(U-B)_{\rm obs}$} \\
	 & ($hh:mm:ss$) & ($\,^{\circ}\,:\,'\,:\,''\,$) & confirmed &  morphology           & counterpart & counterpart  & (erg\,s$^{-1}$) & \multicolumn{2}{c}{({\AA})} & \multicolumn{1}{c}{(mag)} \\ \hline
CALYMHA-S16-5  	&	$10:02:37.01$ & $02:40:28.9$	& NO &	face-on spiral	& \hspace{4pt}--$^\dagger$ 	&	NO 	&	41.44	&	$56$&$22$	&	  0.27$\pm0.07$  \\  
CALYMHA-S16-53 	&	$10:01:54.84$ & $02:16:20.3$	& NO &	face-on spiral	& NO	&	NO 	&	41.47	&	$239$&$137$	&	  0.55  \\  
CALYMHA-S16-61 	&	$10:01:44.50$ & $02:12:34.2$	& NO &	face-on spiral	& NO	&	NO 	&	41.42	&	$216$&$130$	&	  0.26  \\  
CALYMHA-S16-74 	&	$10:01:31.30$ & $02:54:42.8$	& NO &	--   	& -- 	&	NO 	&	41.26	&	$277$&$105$	&	  0.30  \\  
CALYMHA-S16-100	&	$10:00:58.73$ & $02:25:56.3$	& YES &	bright core \& disk 	&	YES    	&	NO 	&	41.88	&	$18$&$1$	&	  0.54  \\  
CALYMHA-S16-149	&	$10:00:13.03$ & $01:51:54.0$	& NO &	disk galaxy   	& NO	&	NO 	&	41.30	&	$74$&$34$	&	  0.55  \\  
CALYMHA-S16-178	&	$09:59:36.48$ & $01:38:45.6$	& NO &	bright core \& disk &	--  	&	NO 	&	41.32	&	$43$&$26$	&	  0.38   \\  
CALYMHA-S16-181	&	$09:59:35.45$ & $01:43:01.6$	& NO &	disk galaxy   	& -- 	&	NO 	&	41.39	&	$79$&$42$	&	  0.33  \\  
CALYMHA-S16-256	&	$09:58:34.68$ & $02:40:37.2$	& NO &	face-on spiral	& -- 	&	NO 	&	41.31	&	$155$&$105$	&	  0.08  \\  
CALYMHA-S16-258	&	$09:58:33.86$ & $02:40:34.7$	& NO &	face-on spiral	& -- 	&	NO 	&	41.33	&	$99$&$61$	&	  0.14   \\  
CALYMHA-S16-265	&	$09:58:26.66$ & $02:28:17.8$	& YES &	bright core \& disk 	&	--  	&	NO 	&	42.12	&	$128$&$8$	&	  0.13  \\  
CALYMHA-S16-275	&	$09:58:13.34$ & $02:05:36.6$	& YES &	bright core \& disk 	&	YES    	&	NO 	&	42.04	&	$27$&$7$	&	  0.26  \\  
CALYMHA-S16-335	&	$09:57:25.01$ & $02:31:19.9$	& NO &	--   	&-- 	&	--  	&	41.18	& $107$&$46$	&	  0.24  \\  
CALYMHA-S16-340	&	$09:57:24.26$ & $02:38:21.1$	& NO &	--   	&-- 	&	NO 	&	41.22	&	$94$&$37$	&	  0.51   \\  
CALYMHA-S16-358	&	$09:56:59.33$ & $01:52:35.8$ 	& NO &	--   	&-- 	&	--  	&	42.20	&	$16$&$2$	&	  0.29  \\  
CALYMHA-S16-388	&	$02:17:17.33$ &	\hspace{-6pt}$-05:14:24.7$	& NO &	 interacting, three nuclei  	&	 -- 	&	--  	&	41.64	&	$52$&$5$	&	 \hspace{-7pt}$-$0.26  \\  
\hline
\end{tabular}}
\begin{tablenotes}
\small
\item $^\dagger$ --: not applicable, there is no coverage in that respective band; 
\end{tablenotes}
\label{tab:indivCII}
\end{threeparttable}
\end{table*}

\begin{table*}
\begin{threeparttable}
\caption{Same as Table~\ref{tab:indivCIII}, but for \C emitters at $z\sim1.05$. A fits version of this table is available online.}
\centering
{\scriptsize
\begin{tabular}{l c c c c c c c r@{$\pm$}l c }
\hline\hline
CALYMHA ID & \multicolumn{1}{c}{RA}   & \multicolumn{1}{c}{DEC} &  $z_{\rm spec}$  &  {\it HST} & {\it Chandra}    & Radio $1.4$ GHz & $\log L_{\rm  CIII]}$ & 
\multicolumn{2}{c}{$EW_{\rm rest}$} & \multicolumn{1}{c}{$(U-B)_{\rm obs}$} \\
	 & ($hh:mm:ss$) & ($\,^{\circ}\,:\,'\,:\,''\,$) & confirmed & morphology            & counterpart & counterpart  & (erg\,s$^{-1}$) & \multicolumn{2}{c}{({\AA})} & \multicolumn{1}{c}{(mag)} \\ \hline
CALYMHA-S16-29 	& $10:02:14.28$ & $02:32:57.8$ &	NO & merger	&	\hspace{4pt}--$^\dagger$ 	&	NO	&	41.87	&	111&51	 & 0.30$\pm0.07$ \\  
CALYMHA-S16-55 	& $10:01:53.71$ & $02:18:58.3$ &	NO &	 disturbed        	&	 NO    	&	NO	&	41.87	&	53&27	&  0.31  \\  
CALYMHA-S16-73 	& $10:01:31.97$ & $02:26:47.4$ &	NO &	 spiral            	&	 NO 	&	NO	&	41.90	&	86&50	&  0.17  \\  
CALYMHA-S16-114	& $10:00:50.83$ & $01:59:55.3$ &	NO &	 disturbed         	&	 NO 	&	NO	&	41.65	&	61&13	&  0.19  \\  
CALYMHA-S16-127	& $10:00:38.50$ & $02:00:04.3$ &	NO &	 interacting     	&	 NO 	&	NO	&	41.69	&	90&20	&  0.15  \\  
CALYMHA-S16-153	& $10:00:09.19$ & $01:54:50.0$ &	NO &	 disturbed         	&	 NO 	&	NO	&	41.77	&	33&15	&  0.29  \\  
CALYMHA-S16-162	& $09:59:54.29$ & $01:39:42.1$ &	NO &	 faint/diffuse    	&	 --  	&	NO	&	41.80	&	179&124	&  0.33  \\  
CALYMHA-S16-182	& $09:59:35.33$ & $01:48:18.4$ &	NO &	 spiral     	&	 NO 	&	NO	&	41.76	&	194&111	&  0.18  \\  
CALYMHA-S16-194	& $09:59:27.79$ & $02:53:13.6$ &	NO &	 --              	&	 -- 	&	 NO	&	41.80	&	77&45	& 0.48  \\  
CALYMHA-S16-205	& $09:59:18.79$ & $01:40:32.2$ &	NO &	diffuse spiral   	&	 -- 	&	 NO	&	41.78	&	37&22	&  0.28  \\  
CALYMHA-S16-212	& $09:59:16.06$ & $01:50:48.1$ &	YES &	 bright core \& disk  	&	 YES   	&	 YES  	&	42.42	&	21&3	&  0.23 \\  
CALYMHA-S16-218	& $09:59:12.84$ & $01:39:09.4$ &	NO &	 interacting galaxies     	&	 -- 	&	 NO	&	41.78	&	96&60	&  0.68  \\  
CALYMHA-S16-237	& $09:59:03.00$ & $01:38:47.4$ &	NO &	 compact galaxy           	&	 -- 	&	 NO	&	42.36	&	73&16	&  0.07 \\  
CALYMHA-S16-262	& $09:58:29.69$ & $02:48:04.3$ &	NO &	 diffuse galaxy           	&	 -- 	&	 NO	&	41.80	&	92&54	&  0.21  \\  
CALYMHA-S16-267	& $09:58:22.82$ & $01:54:55.8$ &	NO &	 disturbed galaxy         	&	 NO	&	NO	&	41.84	&	196&114	&  0.62  \\  
CALYMHA-S16-268	& $09:58:22.30$ & $01:49:13.8$ &	NO &	 diffuse spiral   	&	 -- 	&	  NO	&	41.85	&	144&76	&  0.26  \\  
CALYMHA-S16-273	& $09:58:19.85$ & $02:05:05.6$ &	NO &	 interacting    	&	 NO	&	NO	&	42.02	&	96&65	&  0.36  \\  
CALYMHA-S16-290	& $09:57:51.22$ & $02:35:30.1$ &	NO  &	 faint, compact    	&	 -- 	&	 NO	&	41.69	&	192&99	&  0.04  \\  
CALYMHA-S16-291	& $09:57:51.07$ & $02:21:33.8$ &	NO  & --              	&	 -- 	&	 NO	&	41.68	&	141&66	&  0.39  \\  
CALYMHA-S16-292	& $09:57:50.74$ & $02:21:32.4$ &	NO  & --   	&	 -- 	&	 NO	&	41.59	&	21&9	&  0.21  \\  
CALYMHA-S16-293	& $09:57:50.45$ & $02:28:35.0$ &	NO  &	 diffuse spiral    	&	 -- 	&	 NO	&	41.59	&	87&41	&  0.32  \\  
CALYMHA-S16-297	& $09:57:48.46$ & $02:24:01.4$ &	NO &	 --              	&	 -- 	&	 NO	&	41.61	&	178&103	 & 1.79  \\  
CALYMHA-S16-314	& $09:57:35.57$ & $02:25:50.9$ &	NO  &	 --              	&	 -- 	&	 --   	&	41.63	&	131&65	&  0.29   \\  
CALYMHA-S16-317	& $09:57:33.48$ & $02:26:58.6$ &	NO &	 --              	&	 -- 	&	 --   	&	41.64	&	37&15	 & 0.30  \\  
CALYMHA-S16-326	& $09:57:27.82$ & $02:24:07.9$ &	NO &	 --              	&	 -- 	&	 --   	&	41.70	&	148&68	&  0.05  \\  
CALYMHA-S16-328	& $09:57:26.66$ & $02:24:26.3$ &	NO &	 --              	&	 -- 	&	 --   	&	41.61	&	42&18	&  0.17  \\  
CALYMHA-S16-349	& $09:57:08.86$ & $01:49:52.0$ &	NO &	 --              	&	 -- 	&	 --   	&	41.73	&	44&33	&  0.23  \\  
CALYMHA-S16-352	& $09:57:05.81$ & $01:58:54.5$ &	NO &	 --              	&	 -- 	&	 --   	&	41.69	&	49&38	&  0.07  \\  
CALYMHA-S16-363	& $02:18:17.38$ & \hspace{-6pt}$-04:51:12.2$ & YES & 	 --              	&	 -- 	&	 --   	&	42.96	&	14&1	 & NaN$^\ddagger$ \\  
CALYMHA-S16-371	& $02:17:49.92$ & \hspace{-6pt}$-05:03:15.8$ &	NO &	--              	&	 -- 	&	 --   	&	41.80	&	124&53	 & \hspace{-7pt}$-$0.07  \\  
CALYMHA-S16-381	& $02:17:36.53$ & \hspace{-6pt}$-05:21:56.5$ &	YES &	--              	&	 -- 	&	 --   	&	41.81	&	26&3	 & 0.12  \\  
CALYMHA-S16-385	& $02:17:28.68$ & \hspace{-6pt}$-05:18:49.3$ &	NO &	--              	&	 -- 	&	 --   	&	41.72	&	68&20	 &  \hspace{-7pt}$-$0.30   \\  
CALYMHA-S16-394	& $02:17:13.10$ & \hspace{-6pt}$-05:02:14.6$ &	NO  & --              	&	 -- 	&	 --   	&	41.78	&	69&18	 & \hspace{-7pt}$-$0.22  \\  
CALYMHA-S16-412	& $02:16:41.16$ & \hspace{-6pt}$-05:13:19.2$ &	YES  & bright core, faint arcs	&	 -- 	&	 --   	&	41.87	&	39&5 &	 NaN      \\  \hline
\end{tabular}}
\begin{tablenotes}
\small
\item $^\dagger$ --: not applicable, there is no coverage in that respective band; $^\ddagger$ NaN: not measured because undetected in one band;
\end{tablenotes}
\label{tab:indivCIII}
\end{threeparttable}
\end{table*}

\begin{table*}
\begin{threeparttable}
\caption{Same as Table~\ref{tab:indivCII}, for \CIV emitters at $z\sim1.53$. A fits version of this table can be found online.}
\centering
{\scriptsize
\begin{tabular}{l c c c c c c c r@{$\pm$}l c }
\hline\hline
CALYMHA ID & \multicolumn{1}{c}{RA}   & \multicolumn{1}{c}{DEC} & $z_{\rm spec}$ & {\it HST} & {\it Chandra}    & Radio $1.4$ GHz & $\log L_{\rm  CIV}$ & 
\multicolumn{2}{c}{$EW_{\rm rest}$} & \multicolumn{1}{c}{$(U-B)_{\rm obs}$} \\
	 & ($hh:mm:ss$) & ($\,^{\circ}\,:\,'\,:\,''\,$) & confirmed & morphology           & counterpart & counterpart  & (erg\,s$^{-1}$) & \multicolumn{2}{c}{({\AA})} & \multicolumn{1}{c}{(mag)} \\ \hline
CALYMHA-S16-16	& $10:02:29.16$ &	$02:09:31.7$ & YES	&	\hspace{4pt}p.s.$^\dagger$   &  \hspace{4pt}--$^\ddagger$ 	&   NO  & 42.84	&	8&1	&	0.33$\pm0.07$  \\  
CALYMHA-S16-26	& $10:02:17.98$	&	$01:58:36.8$	& YES &	p.s.   &  --	&   YES  & 43.02	&	66&3	&	    0.56  \\  
CALYMHA-S16-58	& $10:01:47.28$	&	$02:47:29.4$	& YES &	p.s.   &  --	&   NO   	& 42.71	&	41&3	&	  0.28  \\  
CALYMHA-S16-63	& $10:01:41.26$	&	$02:23:08.2$	& YES &	p.s.   & YES	&  NO   	& 42.65	&	97&33	&	  \hspace{-7pt}$-$0.22  \\  
CALYMHA-S16-92	& $10:01:11.98$	&	$02:30:25.2$	& YES &	p.s.   & YES	&  NO   	& 42.59	&	7&1	&	 0.41  \\  
CALYMHA-S16-118  & $10:00:47.95$ &	$02:11:26.9$	& YES &	p.s.	& YES	&  NO  & 42.77	&	12&1	&	 0.64  \\  
CALYMHA-S16-135  & $10:00:28.08$ &	$01:55:47.6$	& YES &	p.s.   &  YES	&  NO   	& 42.88	&	72&4	&	 0.35  \\  
CALYMHA-S16-226  & $09:59:08.69$ &	$02:54:24.5$	& YES &	--    &  -- 	&  possible faint source   & 43.54	&	7&1	&	  0.62  \\
CALYMHA-S16-236  & $09:59:04.01$ &	$02:50:40.6$	& NO &	--    &  -- 	&   NO   	& 42.23	&	138&89	&	 0.49  \\  
CALYMHA-S16-250  & $09:58:48.86$ &	$02:34:41.2$	& YES &	p.s.   &  -- 	&   NO   	& 43.36	&	24&1	&	 0.78  \\  
CALYMHA-S16-274  & $09:58:15.50$ &	$01:49:22.8$	& YES &	p.s.   &  -- 	&  YES  & 43.38	&	16&1	&	 0.14  \\  
CALYMHA-S16-289  & $09:57:52.22$ &	$02:35:24.0$	& NO &	faint p.s.   &  -- 	&  NO   	& 41.99	&	156&96	&	  0.33  \\  
CALYMHA-S16-319  & $09:57:31.70$ &	$02:25:52.3$	& NO &	--    &  -- 	&  --      	& 41.99	&	90&46	&	  0.55  \\  
CALYMHA-S16-325  & $09:57:28.32$ &	$02:25:41.9$	& YES &	--    &  -- 	&  --      	& 43.76	&	31&1	&	  0.48  \\  
CALYMHA-S16-364  & $02:18:16.80$ &	\hspace{-6pt}$-05:00:55.1$	 & YES &	--    &  -- 	&  --      	& 42.19	&	80&28	&	 \hspace{-7pt}$-$0.65 \\  
CALYMHA-S16-372  & $02:17:47.16$ &	\hspace{-6pt}$-04:56:24.7$	& YES &	--    &  -- 	&  --      	& 42.47	&	36&2	&	 0.08   \\
CALYMHA-S16-375  & $02:17:45.22$&	\hspace{-6pt}$-05:01:52.0$ & YES &	--    &  -- 	&  --      	& 42.47	&	19&1	&	 0.43   \\
\hline
\end{tabular}}
\begin{tablenotes}
\small
\item $^\dagger$ p.s.: point source; $^\ddagger$ --: not applicable, there is no coverage in that respective band; 
\end{tablenotes}
\label{tab:indivCIV}
\end{threeparttable}
\end{table*}

\end{document}